\documentclass[journal]{vgtc}                

\ifpdf
  \pdfoutput=1\relax                   
  \usepackage{graphicx}                
  \DeclareGraphicsExtensions{.pdf,.png,.jpg,.jpeg} 
\else
  \usepackage{graphicx}                
  \DeclareGraphicsExtensions{.eps}     
\fi%

\graphicspath{{figures/}{pictures/}{images/}{./}} 

 \usepackage[table,xcdraw]{xcolor}
\definecolor{c1}{rgb}{1,0.54,0.0}
\definecolor{c2}{rgb}{0.4,0,0.2}
\definecolor{c3}{rgb}{0.16, 0.5, 0.0}
\definecolor{c4}{rgb}{0.2 0.2, 1}

\newcommand{\daniel}[1]{{\leavevmode\color{c3}[DS: #1]}}

\newcommand{\revision}[1]{{\leavevmode\color{black}#1}}

\usepackage{microtype}                 
\PassOptionsToPackage{warn}{textcomp}  
\usepackage{textcomp}                  
\usepackage{mathptmx}                  
\usepackage{times}                     
\usepackage{cite}                      
\usepackage{tabu}                      
\usepackage{booktabs}                  
\usepackage{tikz}
\newcommand{\tikzcircle}[2][fill=darkgray]{\tikz[baseline=-0.5ex]\draw[#1,radius=#2] (0,0) circle ;}%

\usepackage{wrapfig}

\usepackage{array}
\newcolumntype{x}[1]{>{\hspace*{-0.1cm}\centering\arraybackslash}p{#1}}

\usepackage[10pt]{moresize}

\usepackage{tabularx}
\usepackage{t1enc}
\usepackage{amssymb}
\usepackage{amsmath}
\usepackage[normalem]{ulem}
\usepackage{cite}
\usepackage{stmaryrd}
\usepackage{dblfloatfix} 
\usepackage{trimclip}
\usepackage{algorithm}
\usepackage{chngcntr}
\usepackage{algpseudocode}
\makeatletter
\DeclareRobustCommand{\shortto}{%
  \mathrel{\mathpalette\short@to\relax}%
}

\newcommand{\short@to}[2]{%
  \mkern2mu
  \clipbox{{.5\width} 0 0 0}{$\m@th#1\vphantom{+}{\shortrightarrow}$}%
}

\onlineid{0}

\vgtccategory{Research}
\vgtcpapertype{application}

\title{MultiSegVA: Using Visual Analytics to Segment \\Biologging Time Series on Multiple Scales}

\author{Philipp Meschenmoser, Juri F. Buchm\"uller, Daniel Seebacher, Martin Wikelski, and Daniel A. Keim.}
\authorfooter{
\item
 Philipp Meschenmoser, Juri F. Buchm\"uller, Daniel Seebacher, and Daniel A. Keim are with the University of Konstanz, Germany. \\ E-mail: \{philipp.meschenmoser, juri.buchmueller, daniel.seebacher, keim\}@uni-konstanz.de 
\item
Martin Wikelski is with the Max Planck Institute of Animal Behavior, Germany. E-mail: wikelski@ab.mpg.de
}

\shortauthortitle{Meschenmoser \MakeLowercase{\textit{et al.}}: Using Visual Analytics to Segment Biologging Time Series on Multiple Scales}
\abstract{
Time series often include patterns and semantics on very different temporal scales. For example, highly resolved biologging time series of migrating animals entail specific patterns and semantics, e.g., migration stages, daily movements, micro-movement frames. However, for segmenting time series, techniques are often applied at a single scale with global parameters, not reflecting the true multi-scale character of the underlying data and models. Thus movement ecologists commonly need cross-domain expertise in statistics to use sophisticated multi-scale segmentation techniques, since there exist no visual-interactive tools that facilitate this kind of analysis.
To close this gap, we present MultiSegVA, a Visual Analytics system for the interactive multi-scale segmentation of unlabeled time series. 
We contribute a novel visual query language to flexibly link numerous techniques on multiple scales to facilitate a visual-interactive analysis. Additionally, we develop a set of segmentation techniques, as well as, suitable visualization in close collaboration with a team of movement ecologists. \daniel{still not happy with this part, we need some more convincing arguments here}	 
We demonstrate the usefulness of MultiSegVA in three use cases in movement ecology and power consumption (...). 
Expert feedback by movement ecologists indicates high domain novelty and making multi-scale methods tangible for more analysts. Additionally, we could show that our approach is applicable to multiple domains, thus laying the foundation for a large range of further research of multi-scale segmentation in other contexts.
}
\abstract{Time series often include patterns and semantics on very different temporal scales. We focus highly resolved biologging time series of animals that entail specific patterns and semantics at, e.g., migration stages, daily movements, micro-movement frames. Though at segmenting time series, techniques are often applied at a single scale with global parameters, not reflecting the multi-scale character. Thus movement ecologists commonly need cross-domain expertise in statistics for more sophisticated multi-scale segmentation techniques, and there are no/few visual-interactive tools that emphasize this kind of analysis. We present MultiSegVA as our main contribution, a Visual Analytics system that provides rich visual-interactive means for facilitating the initial multi-scale segmentation of unlabeled time series. Further, MultiSegVA implements a new visual query language to flexibly link different segmentation techniques on multiple scales. In this publication, we describe and reason visualization as well as interaction design, and specify the visual query language. We demonstrate the applicability and usefulness of MultiSegVA by three use cases from movement ecology and power networks. Preliminary expert feedback by movement ecologists indicates high domain novelty and one way of making multi-scale methods tangible for more analysts. } 
\abstract{High-resolution biologging time series of animals entail specific patterns and semantics at different temporal scales, such as migration stages, days, and micro-movement frames. Segmenting such time series is an essential step before behavior annotation and modeling.  However, segmentation techniques are often applied at a single scale with global parameters and do weakly account for the multi-scale character of the underlying data. More complex multi-scale techniques require careful parameterization and possibly cross-domain expertise, yet there is no visual-interactive tool that strongly supports multi-scale segmentation. To close this gap, we present our MultiSegVA platform, which allows to interactively define segmentation objectives and parameters on multiple scales. MultiSegVA primarily contributes tailored, rich visual-interactive means and VA paradigms for segmenting unlabeled time series on multiple scales. Further, to flexibly compose the multi-scale segmentation, the platform contributes a new visual query language that links a variety of segmentation techniques on multiple scales. Use cases from different domains also require different sets of segmentation techniques catered to the specific needs. To illustrate our approach, we present a domain-oriented set of segmentation techniques developed in collaboration with movement ecologists. We demonstrate the applicability and usefulness of MultiSegVA by two use cases from movement ecology, related to behavior analysis after (a) environment-aware segmentation and (b) progressive clustering. Expert feedback from movement ecologists shows the effectiveness of tailored visual-interactive means at segmenting multi-scale data, enabling them to perform more semantic multi-scale analyses. Moreover we present a third use case from power networks, demonstrating that MultiSegVA is also generalizable to other domains.}

\abstract{Segmenting biologging time series of animals on multiple temporal scales is an essential step that requires complex techniques with careful parameterization and possibly cross-domain expertise. Yet, there is a lack of visual-interactive tools that strongly support such multi-scale segmentation. To close this gap, we present our MultiSegVA platform for interactively defining segmentation techniques and parameters on multiple temporal scales. MultiSegVA primarily contributes tailored, visual-interactive means and visual analytics paradigms for segmenting unlabeled time series on multiple scales. Further, to flexibly compose the multi-scale segmentation, the platform contributes a new visual query language that links a variety of segmentation techniques. To illustrate our approach, we present a domain-oriented set of segmentation techniques derived in collaboration with movement ecologists. We demonstrate the applicability and usefulness of MultiSegVA in two real-world use cases from movement ecology, related to behavior analysis after environment-aware segmentation, and after progressive clustering. Expert feedback from movement ecologists shows the effectiveness of tailored visual-interactive means and visual analytics paradigms at segmenting multi-scale data, enabling them to perform semantically meaningful analyses. A third use case demonstrates that MultiSegVA is generalizable to other domains.}


\CCScatlist{ 
 \CCScat{K.6.1}{Management of Computing and Information Systems}%
{Project and People Management}{Life Cycle};
 \CCScat{K.7.m}{The Computing Profession}{Miscellaneous}{Ethics}
}

\teaser{
  \centering
  \includegraphics[width=0.95\linewidth]{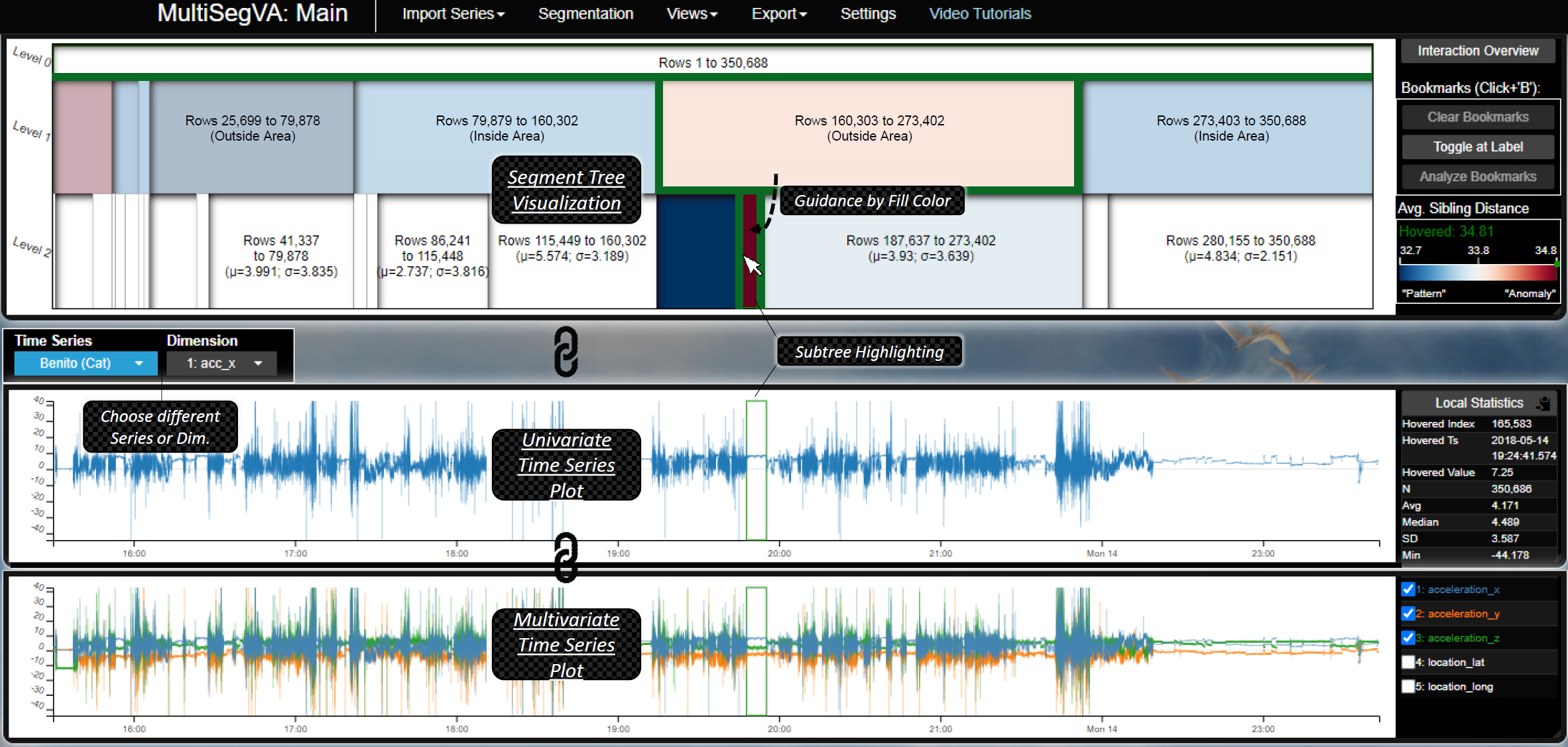}
  \caption{The main window of our MultiSegVA platform for the visual-interactive multi-scale segmentation of biologging time series. After arranging scale-wise segmentation techniques by a new visual query language, the analyst can explore results in tailored visualizations and refine parameters. Here the tree relies on segmentation by geographical area, then recursively by acceleration change points. }
	\label{fig:teaser}
}

\vgtcinsertpkg

\begin{document}


\firstsection{Introduction}
\label{sec:intro2}
\maketitle

Time series often include patterns and semantics on very different temporal scales, but a majority
of segmentation techniques is applied on a single scale with global parameters. More complex multi-scale techniques commonly imply careful parameterization and possibly the need for cross-domain expertise, yet there are no visual-interactive tools with strong support for segmenting time series on multiple scales. We present the MultiSegVA platform that facilitates multi-scale time series segmentation by tailored visual-interactive features, established VA principles, and a new visual query language. In this publication, we focus on biologging time series of moving animals: these time series have prototypical multi-scale character and include widely unexplored behaviors, which are hidden in high resolutions and cardinalities. Additionally, biologging-driven movement ecology is an emerging field \cite{brown2013observing,shepard2008identification,Icarus,Movebank}, triggered by technical advances that enable academia to address open questions in innovative ways.  
The biologging time series stem from miniaturized tags and give high-resolution information about, e.g., an animal's location, tri-axial acceleration, and heart rate. Here, semantics are typically distributed on diverse temporal scales, including life stages, seasons, days, day times, and (micro)movement frames. 
These temporal scales are complemented by spatial scales concerning, e.g., the overall migration range, migration stops, and foraging ranges. 
There are complex scale- and context-specific conditions \cite{benhamou2014scales, thiebault2013splitting, levin1992problem}, implying different energy expenditures, driving factors, and decisions for behavior. Hence, segmenting such time series on a single scale with global parameters does not sufficiently address their multi-scale character.

The relevance of multi-scale segmentation can be further motivated by three reasons. First, analysts can deepen their understanding of how scales relate to each other: e.g., in terms of nesting relations, next to relative scale sizes and types. A multi-scale perspective can even enable ``to gain an insight on an entire knowledge domain or a relevant sub-part'' \cite{nazemi2015semantics}. Second, even without labeled data or thoroughly parameterized single-scale techniques, it is possible to identify fine-grained patterns that are wrapped by lower-scale, context-yielding patterns. Such fine-grained and context-aware patterns are crucial to enrich existing classification and prediction models. Third, demands for more multi-scale analyses originate from domain literature. Such demands can be found in movement ecology and analysis \cite{demvsar2015analysis, levin1992problem, jonsen2013state, andrienko2013movement}, but also in, e.g., medical sciences \cite{alber2019integrating} and social sciences \cite{cash2006scale}. 

However, in practice, segmenting time series on multiple scales is often impeded by several factors. First, multi-scale techniques rely on more in-depth, theoretical foundations and inherent parameters that need to be carefully adapted. Therefore analysts (e.g., movement ecologists) might require cross-domain expertise in statistical multi-scale time series analysis. Second, even with such expertise, it is difficult to decide on scale properties (e.g., size, dimension, number of scales) and further parameters. 
Third, we observe a lack of suitable visual-interactive approaches in related works (Section \ref{sec:relva}) that could strongly support and promote segmenting time series on multiple scales.

To close this gap, we present our web-based \textit{MultiSegVA} platform that allows analysts to visually explore and refine a multi-scale segmentation, which results from a simple way of setting segmentation techniques on multiple scales. In the context of multi-scale segmentation, MultiSegVA primarily contributes the use of tailored visual-interactive features and established VA paradigms (contribution \textbf{C1}). To flexibly configure segmentation techniques and parameters, MultiSegVA includes a new visual query language (VQL, \textbf{C2}) that links a variety of segmentation techniques across multiple scales. These techniques stem in the present case from a set (\textbf{C3}) that was derived together with movement ecologists and covers typical domain use cases. 

Section \ref{sec:tasks} outlines typical tasks and requirements from movement ecology that are contextualized and backed up by the related work Section \ref{sec:relatedwork}. Then Section  \ref{sec:system} shows how we address the first requirements, namely by describing and justifying the platform design. Further requirements are fulfilled by the VQL and its specification in Section \ref{sec:vql}. Three real-world use cases and domain expert feedback from movement ecologists compose Section \ref{sec:eval}, before discussing several aspects.  

In the following, we assume single, discrete, and unlabeled time series, denoted as pair $(t,v)$ and with $n$ records. Let the array
$t=[t[1],\dots, t[n]]$ specify timestamps $\subset \mathbb{R}$ with $\forall{i \in \{2,\dots,n\}}:t[i-1]<t[i]$. Then $v[i]$ is the  (multi-dimensional) value record for timestamp $t[i]$. In general, segmentation shall be the problem of finding a set of index intervals $[\texttt{from}, \texttt{to}] \subset \{1,\dots, n\}$, such that each index $i \in \{1,\dots,n\}$ is included in exactly one of these index intervals.

\section{Domain Background, Tasks, and Requirements}
\label{sec:tasks}
The MultiSegVA platform was developed in close collaboration with domain experts from movement ecology who are working with biologging time series on a day-to-day basis. 

Our domain experts are researchers from the Max Planck Institute of Animal Behavior \revision{that is one of the major driving forces} for the biologging-based, in-depth understanding of animal movements. Hence, the institute coordinates Movebank \cite{Movebank}, an online repository for biologging time series, and the ICARUS initiative \cite{Icarus}, where an ISS antenna gathers signals of global animal movements from space. Likewise, recent advances in biologging technology are opening up a new era in understanding the dynamics and specifics of animal movements. Biologging tags ``have and will become smaller, cheaper and more accessible, new satellite tracking technologies are introduced, data download methodologies become more efficient, battery life increases'' \cite{demvsar2015analysis}.
The domain experts work with resulting time series with GPS and acceleration dimensions, for contributing to subdomains such as animal migration, animal societies, or computational ecology. Such contributions help in better understanding some of today's most urgent issues: e.g., climate change \cite{pecl2017biodiversity, sharma2009impacts}, species decline (cf.  \cite{swan2006removing, sherub2017bio}), disease transmission \cite{han2016global}.
\revision{Thus the movement ecologists are highly experienced in data acquisition, ecological modeling, and corresponding hypothesis testing, while multi-scale time series statistics and advanced machine learning are often less focal points. By interviewing our experts, we could extract the following tasks and related challenges}.

\begin{wrapfigure}{r}{0.24\textwidth}
\vspace{-1.5\baselineskip}
  \begin{center}
    \includegraphics[width=0.24\textwidth]{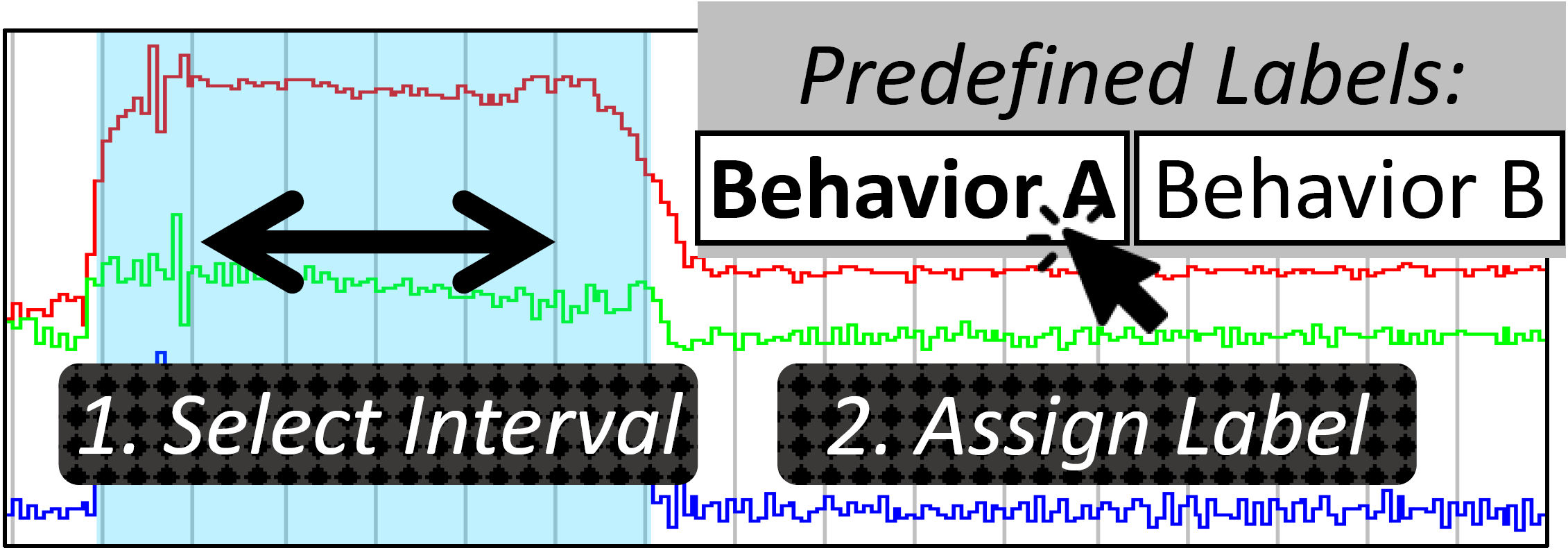}
  \end{center}
  \vspace{-1.6\baselineskip}
  \caption{Manually defining and labeling segments in tri-axial acceleration data.}
  \vspace{-0.7\baselineskip}
  \label{fig:labelassignment}
\end{wrapfigure}

\textbf{Tasks.} \revision{In a typical workflow, our experts mostly follow the KDD pipeline \cite{fayyad1996knowledge}: they begin with sampling, cleaning, standardizing sensor data, before deriving new dimensions (e.g., pitch, ODBA \cite{gleiss2011making}) and plotting GPS as well as acceleration dimensions}. The workflow continues by the main tasks of visual exploration (\textbf{T1}), segmentation (\textbf{T2}), and annotation with behavior labels (\textbf{T3}). \revision{For these tasks, the experts often sequentially apply \textsf{R} and Python scripts, whose adaptation and chaining are tedious. Also, their tools in use do not consider the multi-scale nature of a biologging time series, and they often lack in visual-interactive features, automated techniques, and scalability. Consequently, the existing analysis means cannot sufficiently exploit the true potential and information hidden in the enormous amounts of accruing data \cite{Movebank,Icarus}. For instance, }the experts use a tool that shows tri-axial acceleration data in a zoomable and multivariate time series plot; see the schematic in Figure \ref{fig:labelassignment}. 
The experts interact with the plot, consider contextual data sources, and often try to project themselves into the animal with its motion and movements. To identify and annotate segments of the time series, the experts have to follow a manual process: segments have to be defined by a rectangle selection, before predefined behavior labels are individually assigned. At millions of records, this process can take, as reported, several months. To handle \textbf{T1--T3} while strongly reflecting the multi-scale nature of the time series, the following requirements evolved. 


\textbf{Requirements.} Through regular meetings with the movement ecologists, we derived requirements for the MultiSegVA platform by means of semi-structured interviews. 
Initially, we proposed a set of requirements, based on literature research and supported by interface sketches. These requirements were discussed and refined together with the domain experts. In the following, we describe a high-level set of requirements as result of our discussions. This set complements the requirements in \cite{bernard2016rare} with the aim of satisfying multi-scale aspects.  

\textbf{R1 Integrating Approach.} Insights in time series analysis are typically hidden on different temporal scales\revision{: e.g., daily foraging versus bi-annual migration}. At the same time, each scale comes with scale-specific segmentation objectives, data distributions, patterns, and semantics. \revision{To handle these aspects, the experts now tediously combine techniques from different scripts and tools. For improving the situation, an integrating approach that incorporates various techniques into one scheme is needed.}

\textbf{R2 Multi-Scale Structure \& Specifics.} A temporal scale is typically surrounded by larger temporal scales that have context-yielding patterns, events, anomalies, and semantics. For segmentation, the analyst requires a mean to cover the specifics of surrounding temporal scales and to extract the multi-scale structures of a time series. \revision{Current tools in use support only one temporal scale at a time.}    

\textbf{R3 Flexible Parameterization.}
Time series can be based on several possible multi-scale structures and often the analyst has to explore which structure is most adequate to answer a specific question. \revision{Using scripts for this exploration tends to be inflexible and error-prune as code has to be re-arranged. Hence,} the analyst requires means for the flexible and effortless parameterization of a segmentation task.  

\textbf{R4 Visual-Interactive Features.}
To facilitate configuration and reasoning, the analyst needs extensive visual-interactive features \revision{that clearly go beyond colored line plots}. The methods shall strongly support result exploration, parameter refinement, and segment annotation.

\section{Related Work}
\label{sec:relatedwork}
This section backs up the elicited requirements, contextualizes, and further motivates our work. We initially overview techniques for time series segmentation, before depicting VA approaches for a visual-interactive segmentation process and outlining related VQLs. 
\subsection{Techniques for Time Series Segmentation}
\label{sec:reltechniques}
Many segmentation techniques are applied on a single scale with global parameters and return a simple segment alignment. Keogh et al. present a survey \cite{keogh2004segmenting} on geometric segmentation: with bottom-up, top-down, sliding-window algorithms, next to the efficient hybrid \textit{SWAB}, that try to meet segmentation criteria concerning piecewise linear approximations.

Also change point detection is often applied to split a time series into homogeneous units, see the surveys in \cite{truong2018review, aminikhanghahi2017survey}. The detection is usually steered by a cost function (e.g., maximum-likelihood-based), a search function (e.g., optimal PELT), and additional constraints. With usually more complex foundations in probability theory, Hidden Markov Models (HMMs, \cite{rabiner1989tutorial, zucchini2017hidden, cappe2006inference}), Hidden Semi-Markov Models, and State Space Models (SSMs) are options that tend to capture inherent semantics better than previous techniques.

Next to these generic techniques, there are more domain- or data-specific variants and developments. For example, techniques specifically for (time-annotated) trajectories are depicted in \cite{demvsar2015analysis} and are part of the framework \cite{buchin2011segmenting, buchin2013segmenting} by Buchin et al. Additionally, techniques with relatively novel ways of functioning have recently emerged, e.g., semantic segmentation based on matrix profiles \cite{gharghabi2017matrix}. 

Though, a common aspect of the above techniques is that their default variants do not emphasize the multi-scale character of time series, as they are often applied on a single temporal scale with global parameters.
Such an emphasis can be obtained by more complex variants, e.g., by hierarchical HMMs, multi-level SSMs, or multi-scale change point detection \cite{dette2018multiscale, frick2014multiscale}. So, \cite{cho2012multiscale} depicts a non-parametric Wavelet model to identify change points in periodograms of multiple scales. In general, alternative approaches can emerge from the field of multi-scale time series modeling \cite{ferreira2006multi, ferreira2007multiscale}. Hence to identify multi-scale structures, Vamo{\c{s}} proposes a technique \cite{vamocs2017multiscale} based on a time series' monotony spectrum: i.e., ``the variation of the mean amplitude of the monotonic segments
with respect to the mean local time scale during successive averagings of the time series''\cite{vamocs2017multiscale}. Such techniques have strong theoretical foundations and their overall value is certainly inquestionable. Yet, their increased complexity demands stronger efforts at parameterization 
and possibly cross-domain expertise. With these impeding factors, the next subsection regards to which extent visual-interactive approaches could support the analyst at multi-scale segmentation. 
\subsection{VA Approaches for Time Series Segmentation}
\label{sec:relva}
\begin{table}[h!t]
  \vspace*{-0.15cm}
  \resizebox{\columnwidth}{!}{%
    \begin{tabular}{l|x{0.1cm}|x{0.1cm}|x{0.1cm}|x{0.1cm}|c|c|c}
                                                           & \textbf{R1}                                        & \textbf{R2}                                        & \textbf{R3}                                        & \textbf{R4}                                        & Heterogeneous Data                        & Generalizability                          & Geo-Context                               \\ \toprule \bottomrule
      \textbf{MultiSegVA}                                  & \cellcolor[HTML]{9AB7F6} \tikzcircle{2pt} & \cellcolor[HTML]{9AB7F6} \tikzcircle{2pt} & \cellcolor[HTML]{9AB7F6} \tikzcircle{2pt} & \cellcolor[HTML]{9AB7F6} \tikzcircle{2pt} & \cellcolor[HTML]{9AB7F6} \tikzcircle{2pt} & \cellcolor[HTML]{9AB7F6} \tikzcircle{2pt} & \cellcolor[HTML]{9AB7F6} \tikzcircle{2pt} \\ \toprule \bottomrule
      Alsallakh et al. 2014,~\cite{alsallakh2014visual}    & \cellcolor[HTML]{9AB7F6} \tikzcircle{2pt} & \cellcolor[HTML]{9AB7F6} \tikzcircle{2pt} & \cellcolor[HTML]{9AB7F6} \tikzcircle{2pt} & \cellcolor[HTML]{F2F2F0}                  & \cellcolor[HTML]{F2F2F0}                  & \cellcolor[HTML]{9AB7F6} \tikzcircle{2pt} & \cellcolor[HTML]{F2F2F0}                  \\ \hline
      R\"ohlig et al. 2015,~\cite{rohlig2015parameters}      & \cellcolor[HTML]{9AB7F6} \tikzcircle{2pt} & \cellcolor[HTML]{F2F2F0}                  & \cellcolor[HTML]{9AB7F6} \tikzcircle{2pt} & \cellcolor[HTML]{9AB7F6} \tikzcircle{2pt} & \cellcolor[HTML]{9AB7F6} \tikzcircle{2pt} & \cellcolor[HTML]{9AB7F6} \tikzcircle{2pt} & \cellcolor[HTML]{F2F2F0}                  \\ \hline
      Spretke et al. 2011,~\cite{spretke2011exploration}   &  & \cellcolor[HTML]{F2F2F0}                  & \cellcolor[HTML]{F2F2F0}                  & \cellcolor[HTML]{9AB7F6} \tikzcircle{2pt} & \cellcolor[HTML]{9AB7F6} \tikzcircle{2pt} & \cellcolor[HTML]{F2F2F0}                  & \cellcolor[HTML]{9AB7F6} \tikzcircle{2pt} \\ \hline
      Gao et al. 2013,~\cite{gao2013web}                   & \cellcolor[HTML]{F2F2F0}                  & \cellcolor[HTML]{9AB7F6} \tikzcircle{2pt} & \cellcolor[HTML]{9AB7F6} \tikzcircle{2pt} & \cellcolor[HTML]{F2F2F0}                  & \cellcolor[HTML]{F2F2F0}                  & \cellcolor[HTML]{F2F2F0}                  & \cellcolor[HTML]{F2F2F0}                  \\ \hline
      Zhao et al. 2011,~\cite{zhao2011kronominer}          & \cellcolor[HTML]{F2F2F0}                  & \cellcolor[HTML]{9AB7F6} \tikzcircle{2pt} & \cellcolor[HTML]{F2F2F0}                  & \cellcolor[HTML]{9AB7F6} \tikzcircle{2pt} & \cellcolor[HTML]{9AB7F6} \tikzcircle{2pt} & \cellcolor[HTML]{9AB7F6} \tikzcircle{2pt} & \cellcolor[HTML]{F2F2F0}                  \\ \hline
      Walker et al. 2015,~\cite{walker2015timeclassifier}  & \cellcolor[HTML]{F2F2F0}                  & \cellcolor[HTML]{F2F2F0}                  & \cellcolor[HTML]{F2F2F0}                  & \cellcolor[HTML]{9AB7F6} \tikzcircle{2pt} & \cellcolor[HTML]{F2F2F0}                  & \cellcolor[HTML]{9AB7F6} \tikzcircle{2pt} & \cellcolor[HTML]{F2F2F0}                  \\ \hline
      Bernard et al. 2016,~\cite{bernard2016rare}        & \cellcolor[HTML]{9AB7F6} \tikzcircle{2pt} & \cellcolor[HTML]{F2F2F0}                  & \cellcolor[HTML]{9AB7F6} \tikzcircle{2pt} & \cellcolor[HTML]{9AB7F6} \tikzcircle{2pt} & \cellcolor[HTML]{F2F2F0}                  & \cellcolor[HTML]{F2F2F0}                  & \cellcolor[HTML]{F2F2F0}                  \\ 
    \end{tabular}
  }
  \vspace*{0.15cm}
  \caption{Comparison of related approaches. No related approach fulfills all of the necessary requirements \textbf{R1--R4}. }
  \label{tbl:vacomparison}
  \vspace{-0.7\baselineskip}
\end{table}

The following VA approaches shall enable interactive time series segmentation, where ``humans and machines
cooperate using their respective distinct capabilities for the most effective results'' \cite{keim2008visual}. Several of these approaches are intended for movement and motion analyses. 

Spretke et al. present their VA tool \textit{AnimalEcologyExplorer} \cite{spretke2011exploration}, while emphasizing that VA ``can help to empower the animal tracking community and to foster new insight into the ecology and movement of tracked animals'' \cite{spretke2011exploration}. Their tool is intended for the iterative enrichment and segmentation of biologging time series. The outcomes are communicated in trajectory views, horizon graphs, and line charts, and can be refined by a feedback loop. However, AnimalEcologyExplorer enables only value range segmentation that is applied on a single scale.  

Alsallakh et al. describe an approach \cite{alsallakh2014visual} where the user decides between various segmentation techniques and parameters. The user specifies intervals for their local application, observes outcomes in interactive visualizations, and can follow a feedback loop. In principle, the local application allows segmentation on multiple nested temporal scales, but the proposed one-dimensional color stripe visualization does not promote multi-scale analyses. In contrast, the \textit{KronoMiner}~\cite{zhao2011kronominer} system of Zhao et al. has many visual-interactive features for the multi-scale exploration of segments, which are arranged in a hierarchical circular layout. However, defining these segments is a fully manual task without an automated segmentation technique involved. 

R\"ohlig et al. \cite{rohlig2015parameters} enable the analysis of the impact of different parameter configurations in an interactive prototype.  Here, a time series is segmented multiple times with differently parameterized SSMs. The segmentation results are encoded by aligned color stripes, having several possible color encodings, aggregations, and interaction techniques. A related approach \cite{bernard2016rare} by Bernard et al. aligns color stripes from various segmentation techniques. Though these approaches lack algorithmic and visual multi-scale support. 

Further approaches do not consider typical segmentation techniques inherently, but still help to structure the time series. The \textit{TimeClassifier} tool \cite{walker2015timeclassifier} of Walker et al. structures time series via template matching and visually encodes confidence of matched segments. A tool \cite{gao2013web} of Gao et al. identifies and annotates segments based on a hierarchical support vector machine (SVM). This tool distinguishes between high- and low-level motion patterns but has few visual-interactive features. 

The above VA approaches can be compared to MultiSegVA directly, see Table~\ref{tbl:vacomparison}. Besides the requirements from Section~\ref{sec:tasks}, we also compare support for heterogeneous data and geographical context, as well as generalizability to other domains and diverse kinds of time series. There are several gaps inside the table, and in particular, no related approach fulfills each of the necessary \textbf{R1--R4}. This aspect can be traced back to different overall research objectives. 

Further distinction would be possible by comparing the scalability in terms of temporal resolution, duration, and amount of different time series to be analyzed in context. Yet, these properties are not depicted in all compared works. The range for the amount of displayable time series goes from 1 with high temporal resolution and duration of a few hours \cite{walker2015timeclassifier} to many, but with lower temporal resolution \cite{spretke2011exploration}. We have found no tool that scales to multiple time series with high temporal resolutions ($>1 $ Hz) over durations longer than a few hours. 

To conclude, to the best of the authors' knowledge, there is no VA approach that simultaneously puts a strong emphasis on intelligent multi-scale time series segmentation and the additional support by tailored visualizations and interactions. To fill this gap and address further properties, this publication describes a comprehensive VA system with tailored visual-interactive features and a new visual query language for multi-scale time series segmentation. 
\vspace*{-0.7mm}
\subsection{Related Visual Query Languages}

Visual query languages (VQLs) allow a wide range of users to search databases by replacing textual queries with visual abstractions, thus offloading the complexity of query building from the user to the system. 

Despite many existing VQLs, there is a need for further research at VQLs for temporal domains~\cite{Catarci2009}. The main focus of current research about VQLs for temporal domains lies on event sequences. Decisionflow\cite{gotz2014decisionflow} is a VA solution for high-dimensional temporal event sequence data that provides a VQL to find event sequences by entering preconditions, episodes, and outcomes. A related approach by Kraus et al.~\cite{krause2015supporting} includes the visual-interactive segmentation and filtering of event sequences. 
Another approach, Segmentifier~\cite{dextras2019segmentifier}, iteratively applies filter operations to subdivide click-stream sequence data into smaller segments by imposing constraints for further detail inspection. Though, these aforementioned approaches focus on event sequences and not unlabeled time series data. In contrast, Stein et al.~\cite{stein2019movement}  consider time-annotated trajectories and present a VQL for the hierarchical definition of events. However, their approach only focuses on movement data and does not entail multi-scale segmentation.

%

\section{The MultiSegVA Platform}
\label{sec:system}

With MultiSegVA, the analyst uses tailored visual-interactive features to explore and refine the result of a time series' multi-scale segmentation (\textbf{R2}, \textbf{R4}). This result is a segment tree and stems from a new visual query language (Section \ref{sec:vql}) that enables to specify the hierarchical application order of diverse segmentation techniques. Section \ref{sec:sysoverview} gives an overview on MultiSegVA and its workflow. Section \ref{sec:mainwindow} describes and justifies visualization and interaction choices in the platform's main window, while detail windows are focused in Section \ref{sec:detailwindow}.  

\subsection{System Overview and Feedback-based Workflow}
\label{sec:sysoverview}
\begin{figure}[h]
    \vspace*{-0.3\baselineskip}
\includegraphics[width=\linewidth]{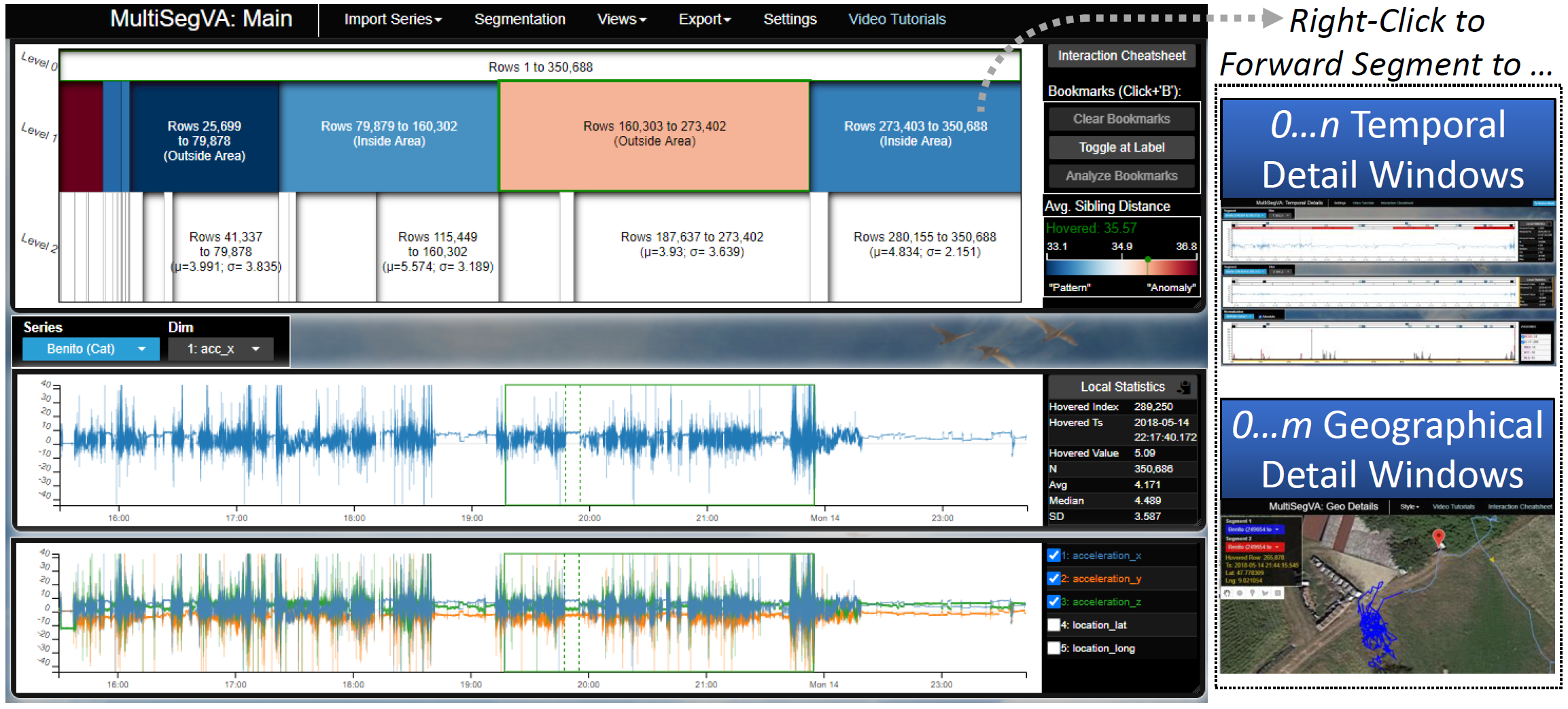}
\caption{The main window (with its segment tree visualization and time series plots) that enables to send segments to $n+m$ detail windows.}
\label{fig:mainwindow}
    \vspace*{-0.3\baselineskip}
\end{figure}

MultiSegVA has one main window where the analyst builds visual queries and analyzes segmentation results inside a hierarchy visualization, closely linked to uni- and multivariate time series plots. The platform includes two additional windows for detail analysis of interesting segments. See Figures \ref{fig:teaser} and \ref{fig:mainwindow} for an overview on this scheme.

 \begin{figure*}[h]
 \centering
  \includegraphics[width=\textwidth]{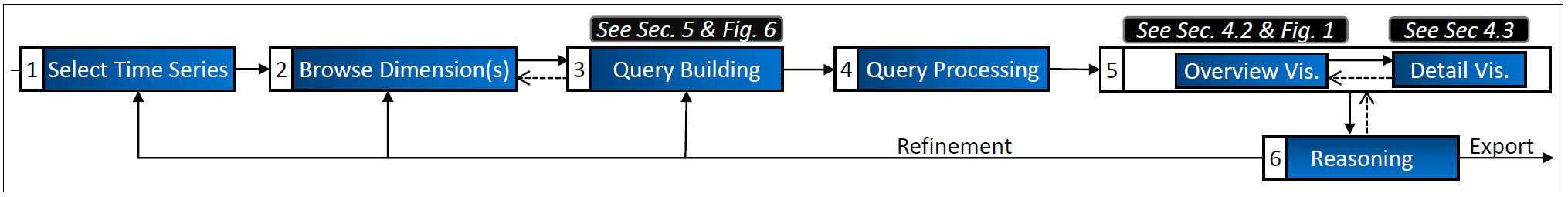}
  \vspace*{-6mm}
  \caption{Feedback-based MultiSegVA workflow, as described in Section \ref{sec:sysoverview}.}
  \label{fig:workflow}
\vspace{-1.5\baselineskip}
\end{figure*} 

 One of these additional windows focuses on temporal details and local anomalies. In presence of geographical dimensions, the analyst can open the other additional window that has an interactive trajectory map and gives the spatial context. MultiSegVA ensures a close linking between these windows, even when the analyst opens them on different computers.            
The intended workflow with MultiSegVA contains the following six steps, also shown in Figure \ref{fig:workflow}. The workflow involves a feedback loop for iterative analysis and refinement.

\textbf{1. Series Selection.} In the main window, the analyst initially decides on a time series: there are four options. \textbf{(A)} The analyst imports a biologging time series from Movebank \cite{Movebank}, which is enabled by a Movebank-MultiSegVA linkage. In principle, the analyst can select a time series out of a set of more than 7,320 studies, with $>2.2$ billion locations in total. \textbf{(B)} The analyst imports an arbitrary time series from a curated set that MultiSegVA offers. \textbf{(C)} Or else, the analyst uploads a .csv file to analyze an own time series. \textbf{(D)} With multiple series imported, it is possible to switch between the series by a select menu.    

\textbf{2. Browse Dimensions.} Now the selected series' first numerical dimension is shown in a univariate time series plot. The plot's y-domain matches to the dimension's minimum and maximum value. To contextualize, all dimensions are shown in a multivariate plot below. At both plots, the analyst can flexibly select the dimension (subset) to show, followed by interactive zooming and panning. Thereby the analyst can develop an initial understanding of dimension-specific properties and can already identify (ir)relevant dimensions and intervals.  

\textbf{3. Query Building.} The analyst arranges segmentation techniques on multiple scales, see the VQL in Section \ref{sec:vql}. Step 2 is helpful again.

\textbf{4. Query Processing.} This step is not steerable for the analyst. 

\textbf{5. Overview + Detail Visualizations.} By a hierarchy visualization, the main window provides an \textit{overview} on the segmentation result, see Figure \ref{fig:teaser} and Section \ref{sec:mainwindow}. This visualization encodes, e.g., a segment's start index, stop index, and segment tree level. By color encoding a similarity measure, the analyst gets guided to interesting segments, indicating possible patterns and anomalies. With a right-click, the analyst sends an interesting segment to the additional windows, tailored for \textit{detail} analysis (Section \ref{sec:detailwindow}). Here the analyst inspects already zoomed-in data values, point-based anomalies, and the spatial context of even very short segments. Simultaneously, the main window shows the \textit{overview}, and steadily enables to select segments for \textit{detail} analysis.    

\textbf{6. Reasoning.} The analyst gets an intuition for scale properties, specific segment lengths and distributions, temporal subtleties, anomalies, spatial context, and corresponding influences. A reasoning step follows, especially in case of domain expertise. Thus in the main window, the analyst can label segments or refine the query. The platform enables such actions by a strong feedback loop: the analyst can dynamically trace back to step 1, 2, 3, or 5. Alternatively, the analyst can export the segment tree to a .csv file and continue externally by own analysis methods or extrapolations.

\textbf{Other Workflow Mappings.} This workflow also has smaller loops, e.g., between steps 2 and 3. Furthermore, like in \cite{alsallakh2014visual}, it is possible to map Keim's VA mantra onto the workflow: ``analyze first - show the important - zoom, filter, analyze further - details on demand'' \cite{keim2008scope}.  Finally, the entire workflow can be also integrated into higher-order workflows, e.g., where the analyst starts with a straightforward segmentation query and iteratively increases the complexity after step 6. 

\subsection{Main Window: Visualization and Interaction}
\label{sec:mainwindow}
Besides the query building dialog, the main window's centerpiece is an icicle visualization \cite{kruskal1983icicle} that encodes the segment tree for a time series with $n$ records. 
While there are additional time series plots, the focus of this section is on describing and justifying the icicle visualization design and then the interaction with it.

\textbf{Visualization.} 
 The icicle visualization recursively partitions display space along the $x$-axis and aligns child stripes below their parent stripe. In MultiSegVA, a stripe's vertical position encodes the respective segment tree level. The stripe height is constant, except for the root stripe that is visually less prominent as it covers all $n$ records. A stripe's $x$-position is given by the linear $xPos(\texttt{from})$ with $xPos:[1,n] \rightarrow [x_{\min}, x_{\max}]$ in the simplest case. It is $xPos(\texttt{from})<xPos(\texttt{from}+1)$, and $xPos(\min(\texttt{to}+1,n))-xPos(\texttt{from})$ yields stripe width. \revision{By using record indices for position and size, a compact view is ensured even for time series with large temporal gaps.} 


The icicle technique is an adequate choice for two reasons. First, the technique provides an intuitive mapping of segment start index and length onto $x$-position and width. Segment start index and length are attributes that asked ecologists emphasized. Second, the icicle technique tends to be space-filling. Here, the icicle technique outperforms competing node-link diagrams that imply often unusable white-space between nodes. While node-link diagrams usually yield less intuitive mappings for a segment's start index and length, these diagrams could be more usable at very large-scale time series.   

To guide the analyst to pattern and anomaly segments, a stripe's fill color encodes a similarity value: the segment's mean sibling distance $\overline{d}$ in a given dimension (set). \revision{Siblings of a segment $a$ with $[\texttt{from},\texttt{to}]$ on level $i$ are those other segments on level $i$ that share with $a$ their parent segment $[\texttt{from}', \texttt{to}']$ on level $i-1$. It is $\texttt{from}'= \min(\texttt{from})$ and $\texttt{to}'= \max(\texttt{to}$). Then MultiSegVA calculates $\overline{d}$ between each segment and its siblings. For this, segments are normalized and then compared by dynamic time warping \cite{berndt1994dtw, rakthanmanon2012searching}.} A relatively low, respectively high,  $\overline{d}$ can indicate a segment-based repeating pattern, respectively anomaly.

A linear color scale $c: D \rightarrow R$ maps $\overline{d}$ to an corresponding fill color. The color scale's range $R$ is given by \textit{ColorBrewer}'s \cite{harrower2003colorbrewer} diverging red-white-blue scheme. A blue stripe can point to a repeating pattern, a red stripe to an anomaly. There are four reasons for choosing this color scheme. \textbf{(1)} The diverging scheme reflects the natural break point $({d}_{min}+\overline{d}_{max})/2$: a whitish stripe indicates a relatively neutral segment. \textbf{(2)} The color scheme can intuitively encode the semantics of $\overline{d}$, especially for anomalies and neutral segments. \textbf{(3)} The colors are well distinguishable, and \textbf{(4)} the scheme is colorblind friendly.

MultiSegVA applies the color-encoding for a hovered stripe and its siblings, while colors for other stripes are faded out. The color scale's domain $D$ is given by the sibling-based minimum and maximum of $\overline{d}$. This choice optimally ensures valid segment comparisons due to relatively similar segment lengths and contextual semantics. Deciding on $D$ was an enduring process where variants were tested and discussed with domain experts: e.g., global color scale domain and mapping, as well as multiple level-wise color scales and legends (see Section \ref{sec:conclusion}).  


Two kinds of labels support understanding and analyzing the icicle visualization. First, labels at the visualization's left denote segment tree levels. Second, labels inside the stripes refer to the start index, stop index, and further annotations. While MultiSegVA derives annotations automatically (e.g., ``inside value range [$r_{\min},r_{\max}$]'', cf. \textbf{T3}), the analyst can also assign own labels. In both cases an uncluttered view is ensured, as a label only appears if it horizontally fits into its stripe. 

\textbf{Interaction.} Next to manual labeling, manifold features for interacting with the icicle visualization are described in a thorough documentation, tutorial videos, and the following lines.

The used zooming and panning behavior ensures quick interaction with the icicle visualization. A left-click moves the clicked stripe into the focus and excludes potentially irrelevant stripes. Then the clicked stripe consumes much space, but the stripe's parent, its direct neighbors, and children are still available for zooming and panning. This behavior involves animated transitions that aid at constructing a mental map and ``maintaining object constancy'' \cite{bederson1999does}. Note that panning is also doable by arrow keys, thus not needing pointing precision at thin stripes. 

Next, hovering over stripes triggers two functions. First, there is subtree highlighting where the hovered stripe and each of its wrapping ancestor stripes obtain a green border. 
Second, hovering links the colored stripes to actual data values: namely, it triggers visual cues inside the aligned time series plots. The visual cues encode start and stop indices of hovered stripes and possible child stripes. Vice versa, hovering the time series plots triggers subtree highlighting. Besides these features, double-clicking a stripe synchronizes the below times series plots to the respective time domain.  

Another interaction feature relates to placing bookmark icons on stripes by a mouse and keyboard interaction. These bookmarks are intended for three uses. \textbf{(1)} Bookmarks enable the icicle's partial growth (Section \ref{subsec:operators}). \textbf{(2)} With bookmarks it is trivial to remember stripes of interest, even after zooming, panning, or segmenting other time series. \textbf{(3)} By an additional button click, the analyst triggers local and ad-hoc anomaly detection for the bookmarked stripes. Section \ref{sec:detailwindow} describes the anomalies' visual mappings in the temporal detail window.

Lastly, right-clicking a stripe triggers that the segment is sent to, and directly focused within, the temporal and geographical detail windows. There the analyst can effortlessly inspect, e.g., zoomed-in values and local anomalies, while retaining the overview in the main window. 

\subsection{Detail Windows: Visualization and Interaction}
\label{sec:detailwindow}
By visiting session-based URLs, the analyst opens additional windows for inspecting (a) temporal details and anomalies, as well as (b) the trajectory and spatial context of a forwarded segment. The scheme enables to retain the overview, extend display space, and do group-based analyses at multiple computers. This subsection describes and justifies the design of first the temporal then geographical detail window. 

\begin{figure}[h!]
\centering
\vspace*{-0.3\baselineskip}
\includegraphics[width=0.9\linewidth]{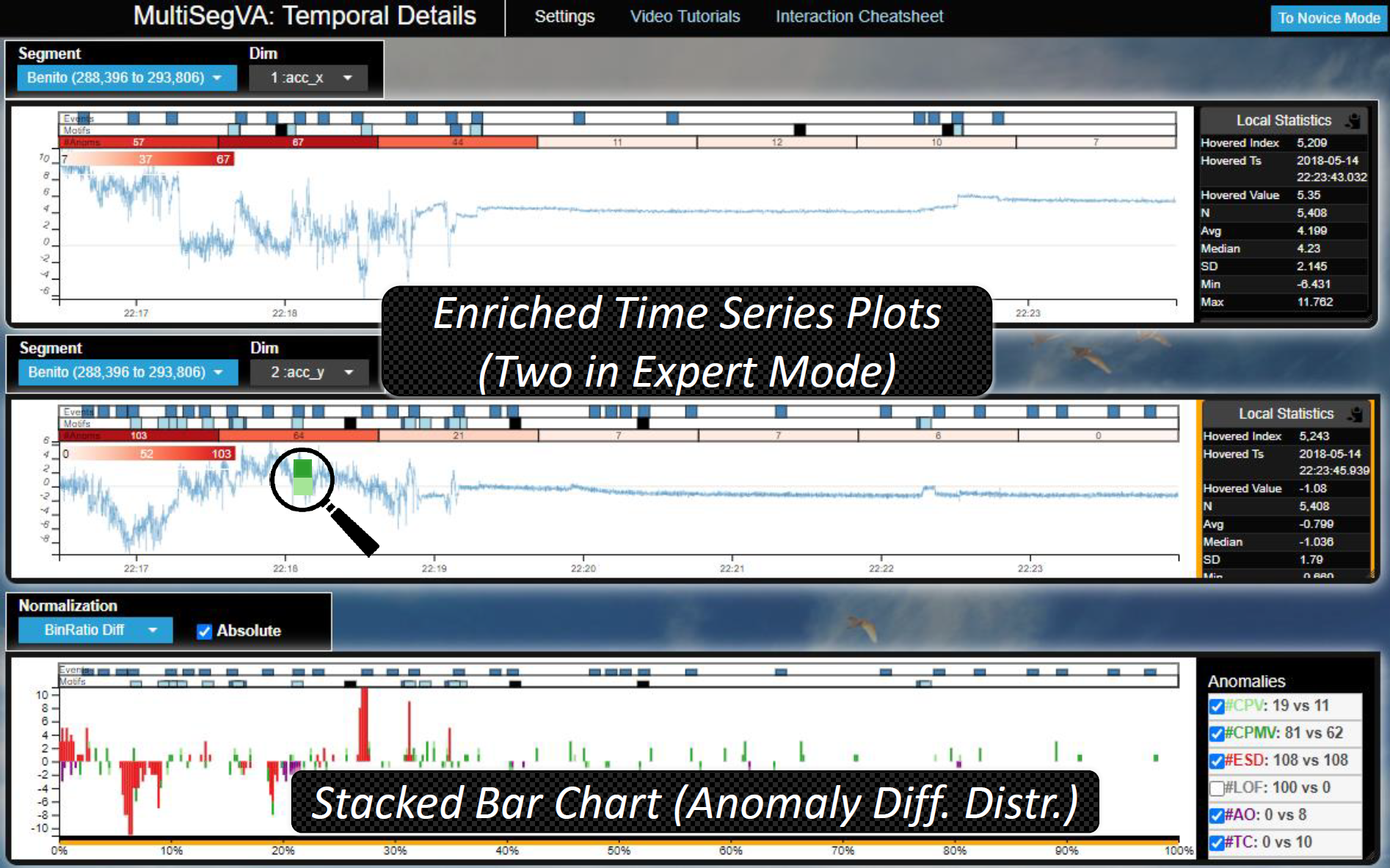}
\caption{The temporal detail window in its expert mode with two enriched time series plots (top, middle) for comparing segments, and a stacked bar chart for showing anomaly distributions in forwarded segments.}
\vspace*{-0.3\baselineskip}
\label{fig:temporaldetailwindow}
\end{figure}

\textbf{Temporal Detail Window.} This window has by default one enriched time series plot and a linked stacked bar chart, both referring to one dimension of a forwarded segment, cf. Figure \ref{fig:temporaldetailwindow}. The analyst can dynamically select previously forwarded segments, respectively dimensions. When the analyst triggers local anomaly detection in the main window, the enriched time series plot shows visual mappings of, e.g., point-based anomalies. The stacked bar chart aggregates these anomalies and shows their distribution over time.

The time series plot is an interactive line chart that offers several interpolation options and can hide line segments at temporal gaps. These properties hold also for the main window's time series plots. In addition, the \textit{enriched} time series plot attaches squares to individual data points, for showing point-based anomalies. A square's categorical fill color encodes the anomaly type: e.g., density-based local outlier \cite{breunig2000lof}, global outlier by \textit{MAD} \cite{leys2013mad} or \textit{S-H-ESD} \cite{hochenbaum2017automatic}, innovative outlier or temporary change by \cite{chen1993joint}. 
The square size is density-dependent to avoid visual overlap. To still guide the viewer to suspicious intervals \revision{and allow their cognitive processing \cite{miller1956magical}}, 7 equal-size rectangles are placed above the line chart. There, ColorBrewer's \textit{reds} scheme encodes the overlaid number of anomalies that are in a $\frac{1}{7}$ interval of the current time domain. Lastly, bars in the top are used for encoding motifs \cite{yeh2016matrix} and events \cite{kang2014detecting}. 

For zooming and panning the plots, the analyst can decide whether the $y$-domain should adjust to local extrema. To improve perception, the lines become thicker if data density is below a threshold. Details (i.e., concrete data values, timestamps) are part of line-snapping tooltips that are density-dependent to avoid flickering effects. Further, the analyst can open an expert mode where two enriched time series plots are aligned. Here it is possible to automatically synchronize time domains and compare two different dimensions of a segment. The zoomed-in time domain is then indicated in the following stacked bar chart.

The stacked bar chart partitions the global time domain of the selected segment into equal-width bins. The bar heights encode normalized anomaly counts per bin $b$ and anomaly type $a$. Again, fill color encodes anomaly type $a$. Beside absolute anomaly counts, various normalizations can address the following questions. Given bin $b$, how many \% of the contained anomalies are of type $a$? Given anomaly type $a$, how many \% of type $a$ anomalies occur in bin $b$? 

For inspecting small bars, the chart allows zooming in the $y$-direction. In $x$-direction, we prevent this interaction to retain the global time domain and yield the context for the (zoomed-in) enriched time series plot(s). By sorting items in an attached legend, the analyst changes the bars' vertical order to inspect another anomaly type's distribution at the baseline. Clicking a bar synchronizes the enriched time series plot(s).  

\textbf{Geographical Detail Window.} In presence of geographical dimensions, the analyst can also forward a segment to the geographical detail window, inspect the directly zoomed-in segment on an interactive map, \revision{and observe the spatial context. Thereby a segment can be enriched by information about, e.g., land use, driving and disturbing factors}. Again, it is at any time possible to select and thereby directly zoom-in two different, previously forwarded, segments. 

A dark blue, respectively red, polyline encodes (lat, long) tuples of a focused segment. For understanding how a focused segment relates to the overall movement, another polyline encodes all (lat, long) tuples of the time series, wrapping the focused segment. This polyline has a lower $z$-index and paler hue to avoid overlap and make it less prominent. To get a better feel for movement direction, polylines are enriched by equidistant arrows that rely on geographical headings. 

Another of these arrows snaps to the (lat$[i]$, long$[i]$) that is closest to the mouse cursor's location. Details for the $i$-th record are to the left of the map. \revision{To ensure a strong linkage to the main window, a map marker is placed at the location of a record that is hovered in previous time series plots. Also, on the interactive map the analyst can use Google Maps utilities to draw markers, circles, rectangles, and custom polygons. Their geometry coordinates can be copied to the clipboard and inserted at the following VQL to segment a time series by a custom geographical area.} 
Finally, it is possible to vary polyline width, toggle map labels, select tile lightness and layout.

\section{VQL for Multi-Scale Segmentation}
\label{sec:vql}
This section answers how to obtain the main window's segment tree.
To model different multi-scale structures and allow flexible parameterization (\textbf{R3}), 
MultiSegVA includes a new visual query language that links various segmentation techniques (\textbf{R1}) across multiple scales. Here the analyst defines the hierarchical application order (\textbf{R2}) of techniques by a simple play with building blocks. So, technique B shall locally operate on the segments resulted from technique A.

\revision{As a major strength, the VQL avoids drawbacks that come with textual queries and code snippets. While the exact multi-scale structure is often unknown a priori, changing the nesting of a text query or code can be error-prone, relatively tedious, and commonly does not entail a compact visual representation. Further strengths (e.g., technique variability, query expressiveness) are depicted within this section.}

\revision{While in principle the VQL can consume any indices-serving technique, we exemplify a domain-oriented set of segmentation techniques that was derived in close collaboration with movement ecologists: see the overview with domain-specific use cases in Table \ref{fig:overview}.} Besides, the VQL also has operators to consider important relations between segments on the same scale. \revision{A multi-scale segmentation query, including techniques and possibly operators, becomes recursively parsed and evaluated in MultiSegVA's Java backend}. The resulting segment tree references a segment's start and stop record index, next to child nodes.
 
Section \ref{subsec:querybuilding} specifies preconditions and the VQL's elementary concept. To make concepts more tangible, the section also depicts query building. Section \ref{subsec:operators} specifies one VQL selector and several operators. 

\subsection{Hierarchical Application and Query Building}
\label{subsec:querybuilding}

\textbf{Preconditions.} An array element $m[i]$ shall describe technique type and parameters (e.g., relevant dimension, $k$)  for applying a segmentation technique, such as from Table \ref{fig:overview}. Then the method call \texttt{getSplitIndices}($m[i]$, $(t,v)$, \texttt{from}, \texttt{to}) shall locally apply $m[i]$ at the index interval [\texttt{from}, \texttt{to}]. The outcome shall be a list $l$ with unique, local, and sorted  split indices $\in \{0,\dots,  \texttt{to}-\texttt{from}+1\} \subset \mathbb N$. The list head shall be padded by $0$ and the list tail by $\texttt{to}-\texttt{from}+1$. 

\textbf{Concept: Hierarchical Application.} The VQL's underlying concept is to iterate over segmentation techniques, and use split indices resulting from $m[i-1]$ to get the application ranges for the next $m[i]$. So, assume that list $l$ contains $a$ split indices of the globally applied technique $m[1]$; it is $[\texttt{from}, \texttt{to}]=[1,t.\texttt{length}]$. To obtain split indices of a subsequent technique $m[2]$,  one calls \texttt{getSplitIndices($m[2]$,\dots)} exactly $(a-1)$ times. Here one iterates over new application ranges, i.e.,  $[\texttt{from}, \texttt{to}]= [l[j]+o$, $l[j+1]+o-1]$. The offset $o$ is relevant for projecting from local to global indices. Obviously one can apply this concept recursively and develop hierarchies with many levels.

This way of specifiying hierarchical application has two benefits. First, technique and dimension variability is ensured. Various segmentation techniques can operate in different dimensions, but the techniques' results are still assimilable into one tree. Second, when segmenting with $m[i]$ there is only one \textit{external} dependence: the application range $[\texttt{from}, \texttt{to}]$ given by $m[i-1]$. Apart from this, $m[i]$ is fully independent of other segmentation steps from any tree level. This independence can be exploited for subquery caching and parallelization. 

\begin{figure}[h]
  \centering
  \vspace*{-0.3\baselineskip}
  \includegraphics[width=0.8\linewidth]{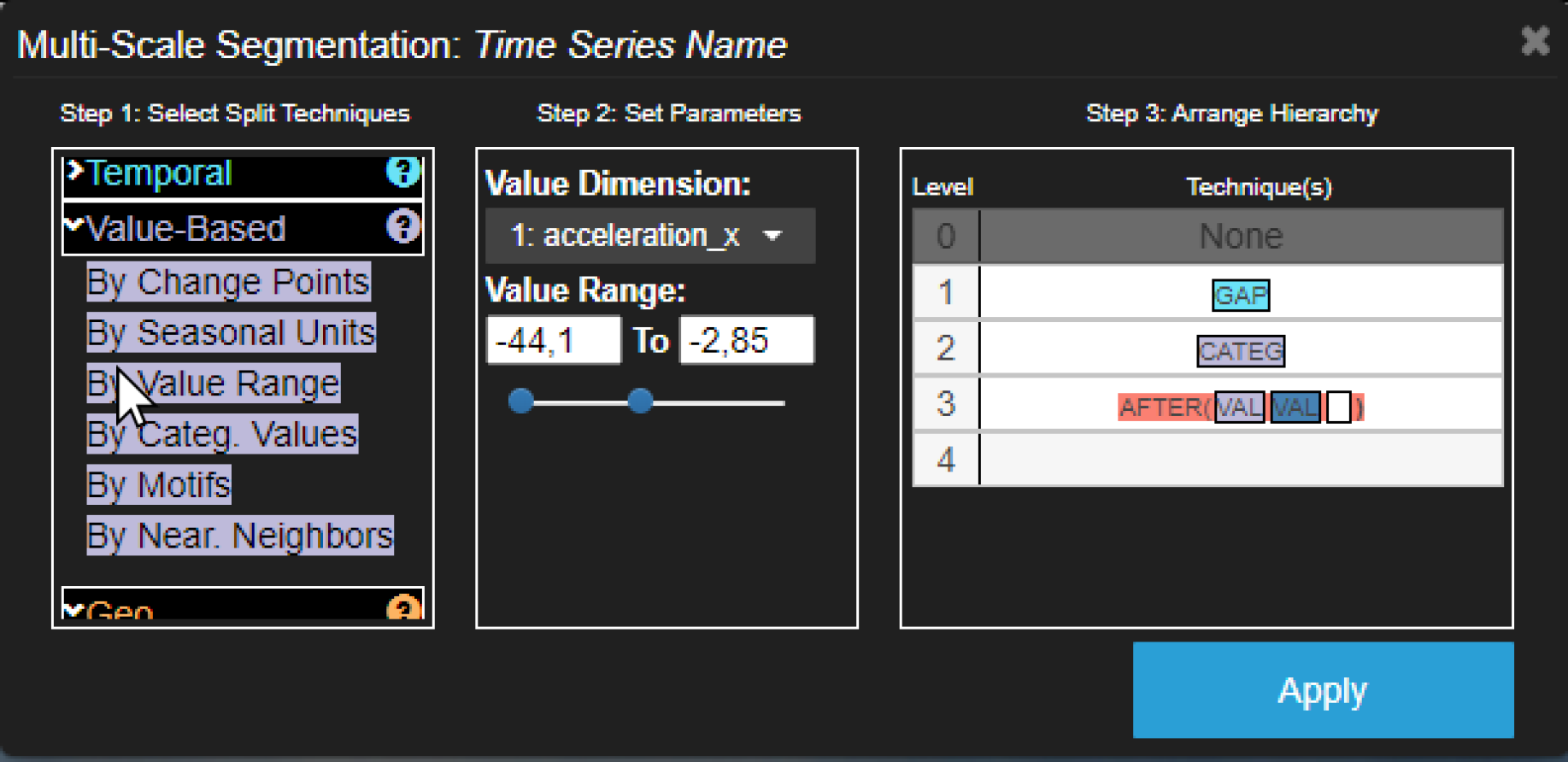}
  
  \caption{The query building dialog is shaped by three steps: choose a segmentation technique, set parameters, arrange the technique/hierarchy. }
    \vspace*{-0.3\baselineskip}
  \label{fig:querybuildingdialog}
\end{figure}

\textbf{Interaction: Query Building.} Before providing further formalization, the remaining subsection shall make the VQL's fundamental concept even more tangible. To this end, Figure \ref{fig:querybuildingdialog} shows the query building dialog. The query in the figure's right segments a time series initially by temporal gaps, then recursively by changes in categorical values, and finally calls a fuzzy operator on tree level 3. Building such a query relies on three simple and effortless interactions. 

\textbf{(1)} Initially the analyst decides on a segmentation technique, see Figure \ref{fig:querybuildingdialog}, left. The techniques are grouped into meaningful categories, and their documentation is linked by question mark icons. \revision{\textbf{(2)} Now the analyst parameterizes the chosen technique. The number of required parameters is kept low, partly given with optional settings and suggested values (e.g., number of motifs). Also, MultiSegVA automatically suggests dimensions with suitable data type for the chosen technique. Longitude and latitude dimensions are internally identified by column headers. Invalid input is caught before backend query processing. \textbf{(3)} Lastly the analyst places the parameterized technique in the query hierarchy (Figure \ref{fig:querybuildingdialog}, right). Here the technique is represented by a building block, whose position is steered by a blinking rectangle. Clicking on an existing building block triggers its blinking and enables to change parameters or replace the technique. Besides, building blocks can be removed via right-click and hierarchy levels are interactively sortable along the $y$-axis.}      

With this base, the specification of further query language features, e.g., of the red \texttt{AFTER} operator in Figure \ref{fig:querybuildingdialog}, follows. 
\subsection{Selector and Operators }
\label{subsec:operators}
This section describes the intuition and relevance of one query language selector and several operators. 

\textbf{Bookmark Selector.} This selector enables a tree's partial growth.
Given $k$ indices from \texttt{getSplitIndices}($m[i-1]$, $(t,v)$, \texttt{from}, \texttt{to}), this selector ensures that the next level's technique $m[i]$ does not perform on each of the $k-1$ new application ranges [\texttt{from}, \texttt{to}]. With this selector it is possible to exclude application ranges that are irrelevant for further analysis, and thereby, to reduce processing times.

\textbf{Operators.} \revision{The VQL has \revision{operators (\texttt{OR}, \texttt{AND}, \texttt{AFTER}, \texttt{NEAR}, \texttt{NOT})} that enable to link techniques also on a \textit{single} tree level (i.e., ``horizontal'' linking).
The operators are implemented by the template in Algorithm \ref{alg:or} that consumes the array $m^{\shortto}$ of techniques to link, as well as the time series and application range. 
The template iterates over $m^{\shortto}$, extracts split indices in $[$\texttt{from}, \texttt{to}$]$, and aggregates them by the operator-dependent function $f$. In the following we focus on the intuition and relevance of \texttt{OR}, \texttt{AND}, \texttt{AFTER}, before referring to further (conceivable) operators.}

\textbf{OR Operator.} This operator merges split indices from multiple techniques and parameterizations. So one could, e.g.,  segment a series by manually specifying multiple geographical areas, apply pattern matching with multiple query patterns (in various dimensions), segment by anomalously high or low value ranges (in various dimensions). \revision{At \texttt{OR}, the template's $f$ is responsible for list concatenation.}  

\setlength{\textfloatsep}{7pt}
\begin{algorithm}
\caption{Operator Template}\label{alg:or}
\begin{algorithmic}[1]
\Procedure{Operator($m^{\shortto}$,$(t,v)$, \textnormal{from}, \textnormal{to})}{}
\State outIndices $\leftarrow$ initEmptyList()
\For{\textbf{each}  $m^{\shortto}[i]$ \textbf{in} $m^{\shortto}$} 
\State newIndices $\leftarrow$ getSplitIndices($m^{\shortto}[i]$, $(t,v)$, from, to)
\State outIndices $\leftarrow$ $f($newIndices, outIndices$)$   
\EndFor
\State \textbf{return} UniqueAndSort(outIndices)
\EndProcedure
\end{algorithmic}
\end{algorithm}

\revision{After this first example, it is evident that an operator description can serve as the first parameter at \texttt{getSplitIndices}(\dots), just like the description for a common technique from Table \ref{fig:overview}}. This property enables nested queries on a single tree level: e.g., a $m^{\shortto}[i]$ could relate to a second \texttt{OR} operator. \revision{Thus highly complex and expressive queries are possible, also with the following template-based operators.} 

\textbf{AND Operator.} This operator retains only these split indices that are returned by all specified techniques in $m^{\shortto}$. Thus this operator is useful for obtaining exact co-occurrences of split indices, given different techniques or parameterizations. \revision{At \texttt{AND}, the template's $f$ corresponds to list intersection.}

\textbf{AFTER Operator.} This operator also focuses on co-occurrences but it is order-dependent and not as restrictive as  \texttt{AND}. Namely, \texttt{AFTER} allows co-occurrences with a temporal tolerance $\theta \geq 0$. Given the techniques $m^{\shortto}[1]$ and $m^{\shortto}[2]$, this operator returns a split index $a$ from $m^{\shortto}[1]$ if there exists a split index $b$ from $m^{\shortto}[2]$ with $b\geq a$ and $b-a \leq \theta$. \revision{Here, $f$ is used to build and extend $\theta$-chains, starting at indices from $m^{\shortto}[1]$ and ending at the indices from the last $m^{\shortto}[i]$.}

\revision{\textbf{Further Operators.} \texttt{NEAR} is similar to \texttt{AFTER} but allows also $b\leq a$. \texttt{NOT} is to be combined with other operators: here, $f$ removes indices from $U=\{1,\dots, \texttt{to}-\texttt{from}\}$. We focused on the integration of above operators to avoid overwhelming interaction choices; still, further (e.g., density-based) operators can be well realized by the template.}

\section{Use Cases and Domain Expert Feedback}
\label{sec:eval}

\revision{MultiSegVA has been developed in close collaboration with movement ecologists and fulfills our audience's requirements (Section \ref{sec:tasks}). We show MultiSegVA's usefulness and applicability by two extensive real-world use cases that were discussed with our domain experts. The use cases originate from live sessions where the domain experts and we were together using MultiSegVA. A third use case from power systems shows the general applicability. Finally, we depict expert feedback that was also used to iteratively improve MultiSegVA.}
\begin{table*}
\centering
    \ssmall
  \begin{tabularx}{\linewidth}{|p{2cm}|p{1.8cm}|p{5cm}|X|}
    \hline
    {\cellcolor[HTML]{20374F} \textcolor{white}{\textbf{Segmentation by}}}  & {\cellcolor[HTML]{20374F} \textcolor{white}{\textbf{Category}}} & {\cellcolor[HTML]{20374F} \textcolor{white}{\textbf{Description/Reference}}}                               & {\cellcolor[HTML]{20374F} \textcolor{white}{\textbf{Movement Ecology Application Cases}}}                                                                                     \\ 
    {\cellcolor[HTML]{C0D7EF} Temporal Gaps}                                & {\cellcolor[HTML]{C0D7EF} Value-Independent}                    & {\cellcolor[HTML]{C0D7EF} Split at anomalous time difference between records.}              & {\cellcolor[HTML]{C0D7EF} Bursts, different recording sessions, changes in sampling rate.}                                                                                            \\ \hline
    {\cellcolor[HTML]{D4EBFF} Bins}                                         & {\cellcolor[HTML]{D4EBFF} Value-Independent}                    & {\cellcolor[HTML]{D4EBFF} Equal or varying depth bins.}                                                    & {\cellcolor[HTML]{D4EBFF} Comparisons, splitting by calendar units, discarding irrelevant segments, adding offsets.}                                                               \\ 
    {\cellcolor[HTML]{C0D7EF} Change Points}                                & {\cellcolor[HTML]{C0D7EF} Generic Value-Based}                  & {\cellcolor[HTML]{C0D7EF} See Section \ref{sec:reltechniques} and \cite{aminikhanghahi2017survey, truong2018review}.}                                                                        & {\cellcolor[HTML]{C0D7EF} Posture changes, unintended collar shifts, simple distinction activity vs. non-activity.}                                                                   \\ \hline
    {\cellcolor[HTML]{D4EBFF} Numerical Value Range}                        & {\cellcolor[HTML]{D4EBFF} Generic Value-Based}                  & {\cellcolor[HTML]{D4EBFF} Split when entering or leaving the value range $\left[r_{\min}, r_{\max} \right]$.}   & {\cellcolor[HTML]{D4EBFF} Flight altitudes or weather conditions, patterns regular in overall shape but elongated.}                                          \\ \hline
    {\cellcolor[HTML]{C0D7EF} Categorical Values}                           & {\cellcolor[HTML]{C0D7EF} Generic Value-Based}                  & {\cellcolor[HTML]{C0D7EF} Split when the categorical value changes.}                                       & {\cellcolor[HTML]{C0D7EF} Validating automated or manual annotations, see the Movebank \cite{Movebank} attribute dictionary.}                                                                   \\ \hline
    {\cellcolor[HTML]{D4EBFF} Seasonal Patterns}                            & {\cellcolor[HTML]{D4EBFF} Generic Value-Based}                  & {\cellcolor[HTML]{D4EBFF} Based on power-maximizing frequency in periodogram.}                             & {\cellcolor[HTML]{D4EBFF} Regular short-time patterns (e.g., flapping), daily fluctuations, latent seasonalities.}                                                           \\ \hline
    {\cellcolor[HTML]{C0D7EF} Motif Representatives}                        & {\cellcolor[HTML]{C0D7EF} Generic Value-Based}                  & {\cellcolor[HTML]{C0D7EF} Split when a motif representative (see \cite{yeh2017multivariate}) begins or ends.}                        & {\cellcolor[HTML]{C0D7EF} Potentially latent recurrences without high assumptions on temporal structure, starting point for further analyses \cite{zhu2016matrix}. Further application cases in \cite{mueen2014motifapplications, meng2008humanmotifs, liu2015motifs}.}           \\ \hline
    {\cellcolor[HTML]{D4EBFF} Pattern Matches}                              & {\cellcolor[HTML]{D4EBFF} Generic Value-Based}                  & {\cellcolor[HTML]{D4EBFF} Split when a matched pattern instance (see \cite{rakthanmanon2012searching}) begins or ends.}                    & {\cellcolor[HTML]{D4EBFF} Follow-up after finding interesting patterns or anomalies, analyzing temporal distribution of matching instances as well as their influence on temporal context.} \\ 
    {\cellcolor[HTML]{C0D7EF} Geographical Area}                                  & {\cellcolor[HTML]{C0D7EF} GPS-Based}                            & {\cellcolor[HTML]{C0D7EF} Split when entering or leaving the specified polygon.}                              & {\cellcolor[HTML]{C0D7EF} A priori estimated home ranges, analyzing relations to specific objects in space, exclude areas with high GPS error.}                                       \\ \hline
    {\cellcolor[HTML]{D4EBFF} Density Clusters}                             & {\cellcolor[HTML]{D4EBFF} GPS-Based}                            & {\cellcolor[HTML]{D4EBFF} Split when entering or leaving a HDBSCAN \cite{campello2013hdbscan} cluster.}                              & {\cellcolor[HTML]{D4EBFF} Automated identification of migration habitats, home ranges, regions with high resource availability. Excluding noise regions.}                             \\ \hline
    {\cellcolor[HTML]{C0D7EF} Local Minima of First-Passage Time}           & {\cellcolor[HTML]{C0D7EF} GPS-Based}                            & {\cellcolor[HTML]{C0D7EF} FPT is ``the time taken by an animal to cross a circle with a given radius'' \cite{pinaud2008fpt}.} & {\cellcolor[HTML]{C0D7EF} Strongly incorporating time component, analyses from ``explorative movements to area-restricted search'' (\cite{ovaskainen2016quantitative}, p. 59).}                                                \\ \hline
  \end{tabularx}
    \vspace*{0.1cm}
  \caption{Domain-oriented set of techniques for structuring biologging time series. Value-independent techniques consider only timestamps or indices. Some techniques are inherently no typical segmentation techniques, still their output can be mapped to the definition in Section 1.}
  \vspace{-0.5cm}
  \label{fig:overview}
\end{table*}
\subsection{Use Case: Environment-Aware Behavior Analysis}
\label{sec:vultures}

\begin{figure}[h!]
\centering
\vspace*{-0.3\baselineskip}
\includegraphics[ width=0.9\linewidth]{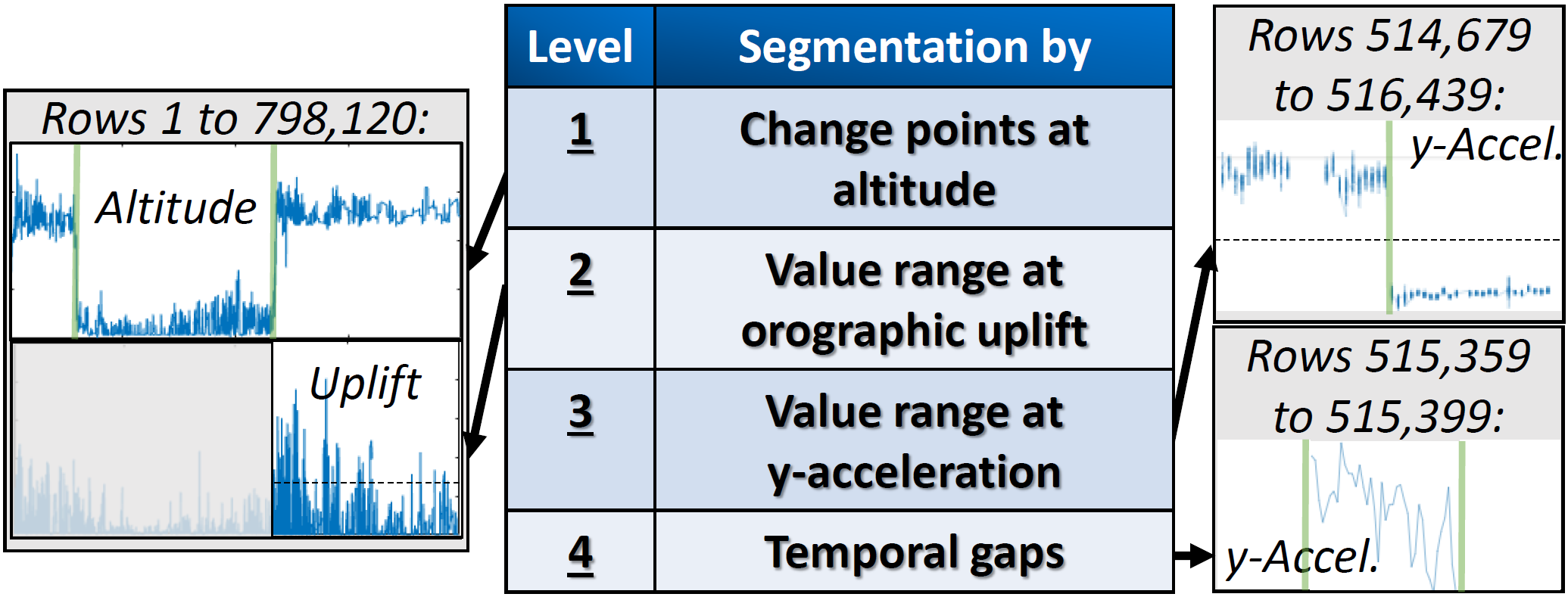}
\vspace*{-0.3\baselineskip}
\caption{From 798,120 to 40 records: the query and scale-wise specifics at our first use case.}
\label{fig:vultureoverview}
\vspace*{-0.3\baselineskip}
\end{figure}

\revision{This use case shows how MultiSegVA helps movement ecologists at identifying environment-aware behaviors in multivariate time series. While behavior is the response to complex environments with many temporal and spatial scales, our experts currently do not have a tool to identify highly specific, environment-aware behaviors. Instead, they apply script chaining that is inflexible at complex environments. Meanwhile, MultiSegVA enables the seamless multi-scale exploration and identification of such behaviors. For this use case, a biologging time series of one migrating Himalayan vulture is focused, including dimensions for location, acceleration, and environmental conditions. See the schematic workflow and scale properties in Figure \ref{fig:vultureoverview}.} 



Himalayan vultures belong to the heaviest flying birds \cite{sherub2016behavioural} and have their main habitat on ``the Roof of the World'', the Tibetan Plateau. Especially juvenile Himalayan vultures perform outstanding altitudinal migrations as part of post-breeding dispersal. Here, the juveniles migrate to warmer regions of Northern India and back to the heights of the Himalaya and Tibetan Plateau. Partly the vultures reach altitudes up to $9,000$ m AMSL. Such altitudinal migrations come with a bandwidth of different environmental conditions: including air densities, wind speeds, temperatures, and resource availability. Movement ecologists are interested in how vultures migrate through these conditions by minimum energy expenditure. Are beneficial environmental conditions always exploited? Are there context-specific differences in acceleration?

The used time series \cite{griffonstudy} stems from Movebank, covers one year, and has $798,120$ records. Each record has two GPS dimensions, one altitude dimension, three acceleration dimensions, and several environmental dimensions. The acceleration data are given on average every 16.16 minutes ($\sigma=$52.19) by a burst of usually 40 records at a frequency of $f$ Hz $ \in \{10.54, 18.74\}$. ``By recording at high resolution [...] but short duration [...] this
strategy aims to sample just one behavior type and avoid behavioral transitions that could complicate automated classification statistics'' \cite{brown2013observing}. Movebank's Env-DATA service \cite{dodge2013envdata} was used to derive environmental dimensions (e.g., orographic uplift).

The analyst imports such a time series by uploading a .csv file. Then in the geographical detail window, the analyst visually explores the trajectory that shows the altitudinal migration from the Tibetan Plateau to India, and vice versa. To distinguish migration stages in $798,120$ records, the analyst starts query building in the main window: change point detection at the altitude dimension. Provided time series plots help in determining the number of change points $k=2$ (i.e., three segments). Visual exploration in the third segment shows that the vulture ascended approximately 5,000 meters within less than four days.

The analyst wants to see how the ascent and high-altitude flight correlate with beneficial environmental conditions. Eastward wind speed, thermal uplift, and in particular orographic uplift have a strong rise during the ascent, shown in the plots and indicating possible migration aids (Figure \ref{fig:vultureoverview}, left). Still, there are intervals with adverse uplift conditions and behaviors on high altitudes. To identify these intervals, the analyst places a bookmark on the high-altitude segment and applies segmentation by the orographic uplift's value range on tree level 2.

As the vulture could also rest on high altitudes, the analyst needs to distinguish between different behaviors, such as fly, rest, or restless. Therefore, segmenting by the $y$-acceleration's value range on level 3 is sufficient, as earlier quantitative experiments could indicate. Adequate value ranges can be determined by visual exploration, some available ground-truth annotations, and domain expertise.

Visual exploration also points out temporal gaps due to burst-wise sampling; thus fully-automated segmentation by temporal gaps is applied to facilitate upcoming analyses. For continuing with statistical tests, the analyst exports the resulted 4-level segment tree to a .csv file. Finally, MultiSegVA returns bursts of flight patterns on high altitudes during adverse weather conditions, and besides, the platform can yield the time series' full multi-scale structures, cf. Figure \ref{fig:lastfig}, a).

MultiSegVA facilitates answering questions as before for several reasons. \textbf{(1)} Without chains of scripts or complex models, MultiSegVA enables high flexibility and multivariate, environment-aware analyses by scale- and dimension-specific segmentation techniques and parameters. \textbf{(2)} Visualizations and interactions are supportive at query building and their scalability is ensured at $798,120$ records.
\textbf{(3)} MultiSegVA has strong I/O abilities and can be integrated into other analysis workflows. 

\subsection{Use Case: Clustering and Behavior Analysis}
\label{sec:cats}
\begin{figure*}[b]
  \vspace*{-0.7\baselineskip}  
  \includegraphics[width=\textwidth]{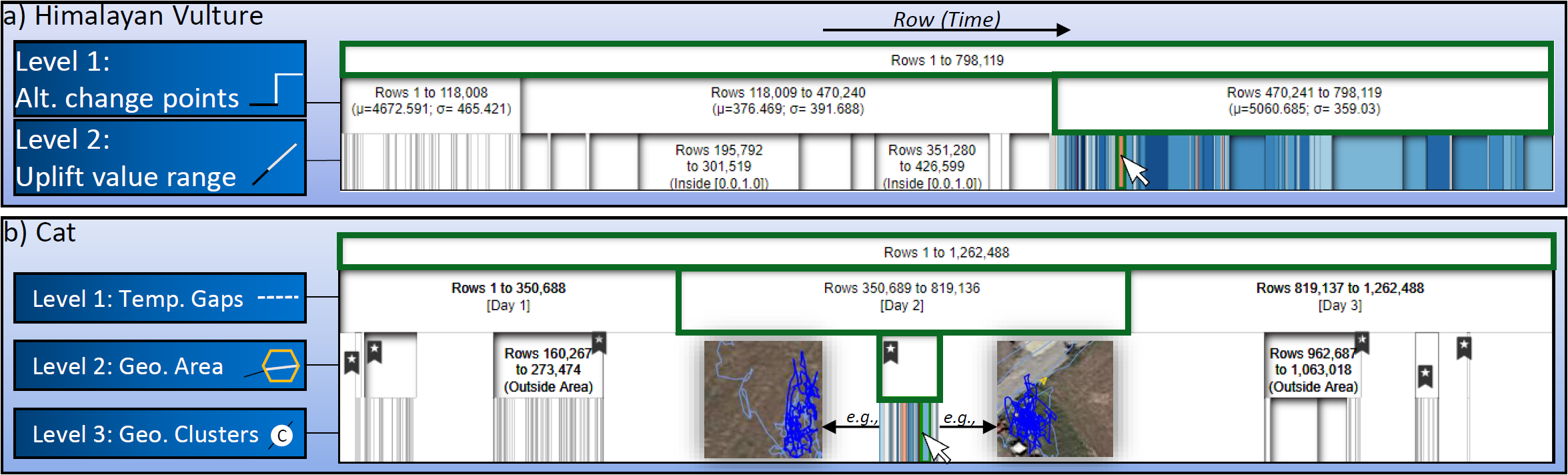}
     \vspace*{-0.75cm}
  \caption{\textbf{a)} Full multi-scale segmentation at splitting the vulture's time series by altitude change points and uplift value range. \textbf{b)} Segmenting the cat's time series by temporal gaps, then locally by area and finally by clusters. Bookmarks allow partial growth and can be automatically attached to labels.}
      \label{fig:lastfig}
\end{figure*}

\revision{For our experts, analyzing spatial density clusters is essential as these regions can point to eating and hence hunting behaviors. Yet, the experts' current analysis means do not enable \textbf{(1)} to flexibly define intervals relevant for the clustering, and \textbf{(2)} to directly apply further intra-cluster analyses. MultiSegVA handles \textbf{(1)} and \textbf{(2)} by progressive clustering: specifically, the analyst recursively chooses relevant intervals on multiple scales, applies one or more spatial clustering steps, and analyzes the intra-cluster acceleration. For this use case, a time series of a free-ranging domestic cat is focused, with data about GPS location and acceleration in more than 1.2 million records. An intermediate outcome of a 3-level segmentation is given by Figure \ref{fig:lastfig}, b).} 

Free-ranging cats have considerable ecological effects on wildlife. For instance, cats interfere in the habitats of wild predators \cite{krauze2012good}, and in particular, the cats' hunting behavior leads to significant reductions of wildlife populations \cite{horn2011home}, e.g., by up to 20.7 billion mammals per year in the US~\cite{loss2013impact}. Yet, so far there is little known about the cats' exact hunting behavior, based on high-resolution biologging data and without disturbing the cats. In fact, identifying detailed hunting and eating behaviors of cats is a very recent topic. Our experts are currently working on this topic and emphasize the value of spatial density clusters. 

While the GPS and acceleration dimensions rely on high sampling rates (continuous 1 Hz and 16 Hz), the time series has recordings from three days that are separated by temporal gaps. Clearly, the three days can reflect entirely different environmental factors and the clusters should not include records from distinctly separated times. As visual exploration suggests, the analyst applies segmentation by temporal gaps on tree level 1. This step extracts the two (out of $>1.2$ million) record indices where the time difference to the next record is anomalous.

Focusing one or more days separately, the analyst now excludes the time series' segments that correspond to irrelevant movements in the trajectory. \revision{So on tree level 2, the analyst applies segmentation by a drawn geographical area (see Table \ref{fig:overview}) to exclude segments inside buildings, which is well feasible at free-ranging cats. This step is useful as there is relatively less prey and the cat is disturbed, but especially, the GPS signal is erroneous inside buildings. 
To additionally reflect segments outside home ranges, it suffices to link these areas via \texttt{OR}.}

Now bookmarking only relevant segments, the analyst locally applies a DBSCAN variant at GPS dimensions on level 3. The local clustering avoids irrelevant data, is faster, and can be again applied on several further levels to better capture hierarchical cluster components.   

Finally, for relevant cluster segments, value range segmentation or change point detection at acceleration dimensions helps in distinguishing coarse behaviors. Exported results can then serve as input at a recent project, i.e., for behavior classifiers with more than 50 classes. To conclude, this use case shares benefits from Section \ref{sec:vultures}, but it also shows how MultiSegVA works within the clustering paradigm and how it unifies more diverse techniques into one compact workflow.

\subsection{Use Case: Power Systems}
To outline MultiSegVA's value even beyond movement ecology, we briefly present 
a use case from the power systems domain. For this, we focus on time series that stem from ENTSO-E \cite{entso} and give country-wide total power loads. The time series are sampled by every 15 minutes and have a time range of several years, up to near realtime. At visual exploration with MultiSegVA, seasonal patterns become evident on various scales: annual (e.g., summer valleys), weekly (e.g., weekend valleys), and daily (e.g., lunch break). To cope with such patterns, MultiSegVA and its VQL already include techniques such as seasonality detection or binning (Table \ref{fig:overview}). MultiSegVA shows the nestedness of seasonal patterns and ensures the structured and guided multi-scale exploration. Also, MultiSegVA provides features for in-depth segment comparisons and local anomaly detection (e.g., Christmas trend) in the temporal detail window. Referring to only two segmentation techniques, this use case already indicates that MultiSegVA variants for other domains are well conceivable. Hereto, specifically tailoring domain-oriented sets of techniques in collaboration with domain experts (e.g., grid operators) is a promising way for most insightful analyses.

    
\subsection{Domain Expert Feedback}
\revision{Iterative feedback by reputable experts in movement ecology allowed us to tailor MultiSegVA for several months and fulfill the elicited requirements.  Now, the use cases become complemented by excerpts of this iterative feedback and summarizing opinions. }

The project began by eliciting requirements, their implementation was reviewed and refined by weekly meetings with the movement ecologists. Here one could discuss relevant segmentation techniques, their parameters, categorization, and application cases in movement ecology. For the main visualization, the ecologists emphasized attributes that appear most important to encode: segment duration, position, and similarity. After prototyping, it was even possible to decide together on the subtleties of visualization and interaction.

After the main development phase, structured live demos of MultiSegVA were given to more than ten movement ecologists who had not been involved until this point. Here we could gather informal feedback that was distinctly positive but still included requests for improvements.

The movement ecologists showed much interest, highlighted the usefulness of given multi-scale segmentation, next to the ease of query building. The query building dialog was denoted as comprehensive and well ordered. The experts liked the close linkage between the segmentation task and visual-interactive features, as well as the linkage between visualizations across computers. Several experts were surprised about MultiSegVA's scalability and short processing times. According to them, MultiSegVA could have saved much time at a recent project. 

However, the movement ecologists depicted the need for more multivariate functions, especially for tri-axial acceleration data. Hence the multivariate plot was added to the main window. While VQL operators already supported multivariate analyses, several segmentation techniques were extended to inherently handle multivariate cases. Anyhow, time series with reduced or projected dimensions can be uploaded. Also, the movement ecologists asked how to know, which interaction features are given and how to use all of them. Thus video tutorials, an overview sheet, and extensive documentation were added. Moreover, the experts asked for export functions that were then added for segment trees and queries. Lastly, the icicle's initially global color scheme appeared difficult to understand, thus leading to the sibling-based scheme. 

After these improvements, three movement ecologists tested MultiSegVA at their own computers without further guidance. Their feedback can give a first intuition on how simple or difficult the independent use is. The domain experts share the opinion that it is straightforward to build queries and segment time series. Still, the experts agree that at least basic prior knowledge about time series segmentation is needed. 

We conclude this section by informal feedback that was gathered throughout the entire project: about (a) possible application cases and (b) perceived novelty. Regarding (a), the movement ecologists see MultiSegVA with its current techniques as a very useful tool for multi-scale exploration and segmentation. One domain expert sees the platform's use for modeling an individual's life history. In general, analyzing populations or with further techniques shall follow after exporting results. One expert would be interested in synchronizing video recordings with the platform, another expert would use MultiSegVA for validation. The experts were also asked (b) to assess the platform's novelty, based on their perspective and domain experiences. Certainly, such opinions depend on the exact background and sub-domain specializations, but still, they give another intuition on the platform's value. The easy segmentation with the emphasis on the multi-scale setting and the strong linkage to visual-interactive methods were seen as most novel. Several movement ecologists agree that MultiSegVA is a progress to what is currently available to them and that it allows them to do more semantically meaningful multi-scale analyses.

\section{Discussion}
\label{sec:conclusion}

MultiSegVA enables comprehensive exploration and refining of multi-scale segmentations by tailored visual-interactive features and VA paradigms. MultiSegVA includes segment tree encoding, subtree highlighting, guidance, density-dependent features, adapted navigation, multi-window support, and a feedback-based workflow. The VQL facilitates exploring and parameterizing different multi-scale structures. Still, few aspects remain for further reflection.  

The icicle visualization meets expert requests and has several benefits. Yet, guiding the user by color to interesting parts of the segment tree is a challenging task. We tested global, level-based, and sibling-based guidance variants and according color fills.  We chose sibling-based guidance (i.e., all siblings of one hovered segment are colored) that optimally captures local similarities, while requiring more navigation effort across levels and nodes. Upcoming works will include an even more effective variant, i.e., guidance to local similarities with little interaction and one fixed color scale. 

Our VQL makes it trivial to \textit{build} a multi-scale segmentation. Query building is a play with building blocks that benefits from strong abstraction and simple interactions. Rather it is difficult to \textit{decide} which multi-scale structure and building blocks are most appropriate: a decision that depends on data, analyst, and tasks. MultiSegVA facilitates this decision by extensive documentation, technique categorization, few technique parameters, and short processing times in a compact workflow. For further support, we plan predefined queries, instant responses at query building, next to parameter and technique suggestions. 

\revision{For suggesting parameters, we will apply estimators \cite{yao1988estimating,chong2001estimating} for the number of change points as well as the elbow method for $knn$-searches. While motif length and HDBSCAN's $minPts$ \cite{campello2013hdbscan} optimally benefit from domain expertise, suggesting other parameters will simplify the interaction and can address another limitation. Now, a technique processes each segment of one scale with the same parameters; thus slight data-dependent parameter modifications will be examined. For technique suggestions, we envision for each technique a scale-wise relevance score that reflects data properties and is part of a rule-based prioritization, shaped by domain expertise and meaningful hierarchies.}      

It is essential to depict the semantics into which MultiSegVA can give insights. First of all, MultiSegVA illuminates diverse multi-scale structures and gives insights on how scales relate to each other. Coarse behaviors can be distinguished by relatively simple techniques, motifs show repetitive behaviors, and $knn$-searches allow the matching with already explored segments. Segment lengths and similarities can be explored, next to local anomalies and spatial contexts. However, with the current techniques it is difficult to broadly capture deeper, behavioral semantics (e.g., chew, scratch). Hereto more complex or learning techniques \revision{(e.g., HMMs, SVMs)} will be needed that neither overfill the interface nor delimit generalizability due to the lack of learned patterns. 

The latter point goes hand in hand with our major limitation and the corresponding implication for upcoming work: integrating even more intelligent methods and automatisms. These plans all relate to aspects from above, i.e., better guidance, technique and more parameter suggestions, as well as techniques for deeper behavioral semantics.   

MultiSegVA relies on requirements by movement ecology experts and stands for an iterative, extensively collaborative and interdisciplinary process.  
We could gather domain feedback on several stages, derive a domain-oriented set of techniques, and even link MultiSegVA to Movebank with $>2.2$ billion animal locations.
With this application domain focused, MultiSegVA underpins the value of multi-scale analyses and is certainly another step forward "to empower the animal tracking community and to foster new insight into the ecology and movement of tracked animals" \cite{spretke2011exploration}.
Meanwhile, our third use case shows that MultiSegVA variants for other domains are conceivable, especially with tailored domain-oriented technique sets. 
This generalizability is promoted by the platform's I/O features and its ability to handle heterogeneous time series with $>1.2$ million records. 
\section{Conclusion}
We presented the web-based MultiSegVA platform that facilitates multi-scale segmentation of biologging time series and enables various semantic analyses. 
To explore results and refine parameters, MultiSegVA primarily contributes the use of visual-interactive features and VA paradigms that are specifically tailored for multi-scale segmentation.
To flexibly model multi-scale segmentations, our VQL is a simple play with building blocks and the second contribution. As an input, the VQL takes techniques out of a domain-oriented set (third contribution) that can fulfill various segmentation objectives, specifically for scale and dimension.
MultiSegVA is shaped by a fruitful collaboration and movement ecology experts see MultiSegVA as a very useful approach for semantically meaningful multi-scale analyses. We are looking forward to integrating even more intelligent methods where interdisciplinary collaboration builds again the basis for effectiveness and quality.


\acknowledgments{
This work has been supported by
the German Research Foundation DFG within Priority Research
Program 1894. We thank the many inspiring researchers from the MPIAB and related institutions who have contributed valuable feedback. They have given us fascinating views into their domain.}

\bibliographystyle{abbrv-doi}

\bibliography{thesis-ref.bib}

\providecommand{\noopsort}[1]{}
\begin{thebibliography}{10}

\bibitem{alber2019integrating}
M.~Alber, A.~B. Tepole, W.~R. Cannon, S.~De, S.~Dura-Bernal, K.~Garikipati,
  G.~Karniadakis, W.~W. Lytton, P.~Perdikaris, L.~Petzold, et~al.
\newblock Integrating machine learning and multiscale modeling - perspectives,
  challenges, and opportunities in the biological, biomedical, and behavioral
  sciences.
\newblock {\em npj Digital Medicine}, 2(1):1--11, 2019.

\bibitem{alsallakh2014visual}
B.~Alsallakh, M.~B{\"o}gl, T.~Gschwandtner, S.~Miksch, B.~Esmael, A.~Arnaout,
  G.~Thonhauser, and P.~Z{\"o}llner.
\newblock A visual analytics approach to segmenting and labeling multivariate
  time series data.
\newblock {\em Proc. of EuroVA}, 14:31--35, 2014.

\bibitem{aminikhanghahi2017survey}
S.~Aminikhanghahi and D.~J. Cook.
\newblock A survey of methods for time series change point detection.
\newblock {\em Knowledge and information systems}, 51(2):339--367, 2017.

\bibitem{andrienko2013movement}
G.~Andrienko, N.~Andrienko, P.~Bak, D.~A. Keim, and S.~Wrobel.
\newblock {\em Visual Analytics of Movement}.
\newblock Springer Berlin Heidelberg, 2013.

\bibitem{bederson1999does}
B.~B. Bederson and A.~Boltman.
\newblock Does animation help users build mental maps of spatial information?
\newblock In {\em Proceedings 1999 IEEE Symposium on Information Visualization
  (InfoVis' 99)}, pp. 28--35. IEEE, 1999.

\bibitem{benhamou2014scales}
S.~Benhamou.
\newblock Of scales and stationarity in animal movements.
\newblock {\em Ecology letters}, 17(3):261--272, 2014.

\bibitem{bernard2016rare}
J.~Bernard, E.~Dobermann, M.~B{\"o}gl, M.~R{\"o}hlig, A.~V{\"o}gele, and
  J.~Kohlhammer.
\newblock Visual-interactive segmentation of multivariate time series.
\newblock In {\em EuroVis Workshop on Visual Analytics (EuroVA). Eurographics},
  2016.

\bibitem{berndt1994dtw}
D.~J. Berndt and J.~Clifford.
\newblock Using dynamic time warping to find patterns in time series.
\newblock In {\em KDD workshop}, vol.~10, pp. 359--370. Seattle, WA, 1994.

\bibitem{breunig2000lof}
M.~M. Breunig, H.-P. Kriegel, R.~T. Ng, and J.~Sander.
\newblock Lof: identifying density-based local outliers.
\newblock In {\em ACM sigmod record}, vol.~29, pp. 93--104. ACM, 2000.

\bibitem{brown2013observing}
D.~D. Brown, R.~Kays, M.~Wikelski, R.~Wilson, and A.~P. Klimley.
\newblock Observing the unwatchable through acceleration logging of animal
  behavior.
\newblock {\em Animal Biotelemetry}, 1(1):20, 2013.

\bibitem{buchin2011segmenting}
M.~Buchin, A.~Driemel, M.~Van~Kreveld, and V.~Sacrist{\'a}n.
\newblock Segmenting trajectories: A framework and algorithms using
  spatiotemporal criteria.
\newblock {\em Journal of Spatial Information Science}, 2011(3):33--63, 2011.

\bibitem{buchin2013segmenting}
M.~Buchin, H.~Kruckenberg, and A.~K{\"o}lzsch.
\newblock Segmenting trajectories by movement states.
\newblock In {\em Advances in spatial data handling}, pp. 15--25. Springer,
  2013.

\bibitem{campello2013hdbscan}
R.~J. G.~B. Campello, D.~Moulavi, and J.~Sander.
\newblock Density-based clustering based on hierarchical density estimates.
\newblock In {\em Pacific-Asia conference on knowledge discovery and data
  mining}, pp. 160--172. Springer, 2013.

\bibitem{cappe2006inference}
O.~Capp{\'e}, E.~Moulines, and T.~Ryden.
\newblock {\em Inference in Hidden Markov Models}.
\newblock Springer Series in Statistics. Springer New York, 2006.

\bibitem{cash2006scale}
D.~W. Cash, W.~N. Adger, F.~Berkes, P.~Garden, L.~Lebel, P.~Olsson,
  L.~Pritchard, and O.~Young.
\newblock Scale and cross-scale dynamics: governance and information in a
  multilevel world.
\newblock {\em Ecology and society}, 11(2), 2006.

\bibitem{Catarci2009}
T.~Catarci.
\newblock {\em Visual Query Language}, pp. 3399--3405.
\newblock Springer US, Boston, MA, 2009. doi: {{%
10\hspace{.1pt}\discretionary{.}{%
}{.}\hspace{.4pt}1007\discretionary{/}{%
}{/}978\discretionary{%
}{-}{-}0\discretionary{%
}{-}{-}387\discretionary{%
}{-}{-}39940\discretionary{%
}{-}{-}9\_448}}


\bibitem{chen1993joint}
C.~Chen and L.-M. Liu.
\newblock Joint estimation of model parameters and outlier effects in time
  series.
\newblock {\em Journal of the American Statistical Association},
  88(421):284--297, 1993.

\bibitem{cho2012multiscale}
H.~Cho and P.~Fryzlewicz.
\newblock Multiscale and multilevel technique for consistent segmentation of
  nonstationary time series.
\newblock {\em Statistica Sinica}, pp. 207--229, 2012.

\bibitem{chong2001estimating}
T.~T.-L. Chong.
\newblock Estimating the locations and number of change points by the
  sample-splitting method.
\newblock {\em Statistical papers}, 42(1):53--79, 2001.

\bibitem{demvsar2015analysis}
U.~Dem{\v{s}}ar, K.~Buchin, F.~Cagnacci, K.~Safi, B.~Speckmann, N.~Van~de
  Weghe, D.~Weiskopf, and R.~Weibel.
\newblock Analysis and visualisation of movement: an interdisciplinary review.
\newblock {\em Movement Ecology}, 3(1):5, 2015.

\bibitem{dette2018multiscale}
H.~Dette, T.~Sch{\"u}ler, and M.~Vetter.
\newblock Multiscale change point detection for dependent data.
\newblock {\em arXiv preprint arXiv:1811.05956}, 2018.

\bibitem{dextras2019segmentifier}
K.~Dextras-Romagnino and T.~Munzner.
\newblock Segmentifier: Interactive refinement of clickstream data.
\newblock In {\em Computer Graphics Forum}, vol.~38, pp. 623--634. Wiley Online
  Library, 2019.

\bibitem{dodge2013envdata}
S.~Dodge, G.~Bohrer, R.~Weinzierl, S.~C. Davidson, R.~Kays, D.~Douglas,
  S.~Cruz, J.~Han, D.~Brandes, and M.~Wikelski.
\newblock The environmental-data automated track annotation (env-data) system:
  linking animal tracks with environmental data.
\newblock {\em Movement Ecology}, 1(1):3, 2013.

\bibitem{fayyad1996knowledge}
U.~M. Fayyad, G.~Piatetsky-Shapiro, P.~Smyth, et~al.
\newblock Knowledge discovery and data mining: Towards a unifying framework.
\newblock In {\em KDD}, vol.~96, pp. 82--88, 1996.

\bibitem{ferreira2007multiscale}
M.~A. Ferreira, A.~Marco, and H.~K. Lee.
\newblock {\em Multiscale modeling: a Bayesian perspective}.
\newblock Springer Science \& Business Media, 2007.

\bibitem{ferreira2006multi}
M.~A. Ferreira, M.~West, H.~K. Lee, D.~M. Higdon, et~al.
\newblock Multi-scale and hidden resolution time series models.
\newblock {\em Bayesian Analysis}, 1(4):947--967, 2006.

\bibitem{frick2014multiscale}
K.~Frick, A.~Munk, and H.~Sieling.
\newblock Multiscale change point inference.
\newblock {\em Journal of the Royal Statistical Society: Series B (Statistical
  Methodology)}, 76(3):495--580, 2014.

\bibitem{gao2013web}
L.~Gao, H.~A. Campbell, O.~R. Bidder, and J.~Hunter.
\newblock A web-based semantic tagging and activity recognition system for
  species' accelerometry data.
\newblock {\em Ecological Informatics}, 13:47--56, 2013.

\bibitem{gharghabi2017matrix}
S.~Gharghabi, Y.~Ding, C.-C.~M. Yeh, K.~Kamgar, L.~Ulanova, and E.~Keogh.
\newblock Matrix profile viii: domain agnostic online semantic segmentation at
  superhuman performance levels.
\newblock In {\em 2017 IEEE International Conference on Data Mining (ICDM)},
  pp. 117--126. IEEE, 2017.

\bibitem{gleiss2011making}
A.~C. Gleiss, R.~P. Wilson, and E.~L. Shepard.
\newblock Making overall dynamic body acceleration work: on the theory of
  acceleration as a proxy for energy expenditure.
\newblock {\em Methods in Ecology and Evolution}, 2(1):23--33, 2011.

\bibitem{gotz2014decisionflow}
D.~Gotz and H.~Stavropoulos.
\newblock Decisionflow: Visual analytics for high-dimensional temporal event
  sequence data.
\newblock {\em IEEE transactions on visualization and computer graphics},
  20(12):1783--1792, 2014.

\bibitem{han2016global}
B.~A. Han, A.~M. Kramer, and J.~M. Drake.
\newblock Global patterns of zoonotic disease in mammals.
\newblock {\em Trends in parasitology}, 32(7):565--577, 2016.

\bibitem{harrower2003colorbrewer}
M.~Harrower and C.~A. Brewer.
\newblock Colorbrewer.org: an online tool for selecting colour schemes for
  maps.
\newblock {\em The Cartographic Journal}, 40(1):27--37, 2003.

\bibitem{hochenbaum2017automatic}
J.~Hochenbaum, O.~S. Vallis, and A.~Kejariwal.
\newblock Automatic anomaly detection in the cloud via statistical learning.
\newblock {\em arXiv preprint arXiv:1704.07706}, 2017.

\bibitem{horn2011home}
J.~A. Horn, N.~Mateus-Pinilla, R.~E. Warner, and E.~J. Heske.
\newblock Home range, habitat use, and activity patterns of free-roaming
  domestic cats.
\newblock {\em The Journal of Wildlife Management}, 75(5):1177--1185, 2011.

\bibitem{jonsen2013state}
I.~Jonsen, M.~Basson, S.~Bestley, M.~V. Bravington, T.~Patterson, M.~W.
  Pedersen, R.~B. Thomson, U.~H. Thygesen, and S.~J. Wotherspoon.
\newblock State-space models for bio-loggers: A methodological road map.
\newblock {\em Deep Sea Research Part II: Topical Studies in Oceanography},
  88:34--46, 2013.

\bibitem{kang2014detecting}
Y.~Kang, D.~Belu{\v{s}}i{\'c}, and K.~Smith-Miles.
\newblock Detecting and classifying events in noisy time series.
\newblock {\em Journal of the Atmospheric Sciences}, 71(3):1090--1104, 2014.

\bibitem{keim2008visual}
D.~A. Keim, G.~Andrienko, J.-D. Fekete, C.~G{\"o}rg, J.~Kohlhammer, and
  G.~Melan{\c{c}}on.
\newblock Visual analytics: Definition, process, and challenges.
\newblock In {\em Information visualization}, pp. 154--175. Springer, 2008.

\bibitem{keim2008scope}
D.~A. Keim, F.~Mansmann, J.~Schneidewind, J.~Thomas, and H.~Ziegler.
\newblock Visual analytics: Scope and challenges.
\newblock In {\em Visual data mining}, pp. 76--90. Springer, 2008.

\bibitem{keogh2004segmenting}
E.~Keogh, S.~Chu, D.~Hart, and M.~Pazzani.
\newblock Segmenting time series: A survey and novel approach.
\newblock In {\em Data mining in time series databases}, pp. 1--21. World
  Scientific, 2004.

\bibitem{krause2015supporting}
J.~Krause, A.~Perer, and H.~Stavropoulos.
\newblock Supporting iterative cohort construction with visual temporal
  queries.
\newblock {\em IEEE transactions on visualization and computer graphics},
  22(1):91--100, 2015.

\bibitem{krauze2012good}
D.~Krauze-Gryz, J.~Gryz, J.~Goszczy{\'n}ski, P.~Chylarecki, and M.~Zmihorski.
\newblock The good, the bad, and the ugly: space use and intraguild
  interactions among three opportunistic predators - cat (felis catus), dog
  (canis lupus familiaris), and red fox (vulpes vulpes) - under human pressure.
\newblock {\em Canadian Journal of Zoology}, 90(12):1402--1413, 2012.

\bibitem{kruskal1983icicle}
J.~B. Kruskal and J.~M. Landwehr.
\newblock Icicle plots: Better displays for hierarchical clustering.
\newblock {\em The American Statistician}, 37(2):162--168, 1983.

\bibitem{levin1992problem}
S.~A. Levin.
\newblock The problem of pattern and scale in ecology: the robert h. macarthur
  award lecture.
\newblock {\em Ecology}, 73(6):1943--1967, 1992.

\bibitem{leys2013mad}
C.~Leys, C.~Ley, O.~Klein, P.~Bernard, and L.~Licata.
\newblock Detecting outliers: Do not use standard deviation around the mean,
  use absolute deviation around the median.
\newblock {\em Journal of Experimental Social Psychology}, 49(4):764--766,
  2013.

\bibitem{liu2015motifs}
B.~Liu, J.~Li, C.~Chen, W.~Tan, Q.~Chen, and M.~C. Zhou.
\newblock Efficient motif discovery for large-scale time series in healthcare.
\newblock {\em IEEE Transactions on Industrial Informatics}, 11(3):583--590,
  2015.

\bibitem{loss2013impact}
S.~R. Loss, T.~Will, and P.~P. Marra.
\newblock The impact of free-ranging domestic cats on wildlife of the united
  states.
\newblock {\em Nature communications}, 4(1):1--8, 2013.

\bibitem{meng2008humanmotifs}
J.~Meng, J.~Yuan, M.~Hans, and Y.~Wu.
\newblock Mining motifs from human motion.
\newblock In {\em Eurographics (Short Papers)}, pp. 71--74, 2008.

\bibitem{miller1956magical}
G.~A. Miller.
\newblock The magical number seven, plus or minus two: Some limits on our
  capacity for processing information.
\newblock {\em Psychological review}, 63(2):81, 1956.

\bibitem{mueen2014motifapplications}
A.~Mueen.
\newblock Time series motif discovery: dimensions and applications.
\newblock {\em Wiley Interdisciplinary Reviews: Data Mining and Knowledge
  Discovery}, 4(2):152--159, 2014.

\bibitem{nazemi2015semantics}
K.~Nazemi, D.~Burkhardt, E.~Ginters, and J.~Kohlhammer.
\newblock Semantics visualization--definition, approaches and challenges.
\newblock {\em Procedia Computer Science}, 75:75--83, 2015.

\bibitem{ovaskainen2016quantitative}
O.~Ovaskainen, H.~J. de~Knegt, and M.~M. Delgado.
\newblock {\em Quantitative Ecology and Evolutionary Biology: Integrating
  models with data}.
\newblock Oxford Series in Ecology and Evolution. OUP Oxford, 2016.

\bibitem{pecl2017biodiversity}
G.~T. Pecl, M.~B. Ara{\'u}jo, J.~D. Bell, J.~Blanchard, T.~C. Bonebrake, I.-C.
  Chen, T.~D. Clark, R.~K. Colwell, F.~Danielsen, B.~Eveng{\aa}rd, et~al.
\newblock Biodiversity redistribution under climate change: Impacts on
  ecosystems and human well-being.
\newblock {\em Science}, 355(6332):eaai9214, 2017.

\bibitem{pinaud2008fpt}
D.~Pinaud.
\newblock Quantifying search effort of moving animals at several spatial scales
  using first-passage time analysis: effect of the structure of environment and
  tracking systems.
\newblock {\em Journal of Applied Ecology}, 45(1):91--99, 2008.

\bibitem{rabiner1989tutorial}
L.~R. Rabiner.
\newblock A tutorial on hidden markov models and selected applications in
  speech recognition.
\newblock {\em Proceedings of the IEEE}, 77(2):257--286, 1989.

\bibitem{rakthanmanon2012searching}
T.~Rakthanmanon, B.~Campana, A.~Mueen, G.~Batista, B.~Westover, Q.~Zhu,
  J.~Zakaria, and E.~Keogh.
\newblock Searching and mining trillions of time series subsequences under
  dynamic time warping.
\newblock In {\em Proceedings of the 18th ACM SIGKDD international conference
  on Knowledge discovery and data mining}, pp. 262--270. ACM, 2012.

\bibitem{rohlig2015parameters}
M.~R{\"o}hlig, M.~Luboschik, F.~Kr{\"u}ger, T.~Kirste, H.~Schumann,
  M.~B{\"o}gl, B.~Alsallakh, and S.~Miksch.
\newblock Supporting activity recognition by visual analytics.
\newblock In {\em Visual Analytics Science and Technology (VAST), 2015 IEEE
  Conference on}, pp. 41--48. IEEE, 2015.

\bibitem{sharma2009impacts}
S.~Sharma, S.~Couturier, and S.~D. Cote.
\newblock Impacts of climate change on the seasonal distribution of migratory
  caribou.
\newblock {\em Global Change Biology}, 15(10):2549--2562, 2009.

\bibitem{shepard2008identification}
E.~L. Shepard, R.~P. Wilson, F.~Quintana, A.~G. Laich, N.~Liebsch, D.~A.
  Albareda, L.~G. Halsey, A.~Gleiss, D.~T. Morgan, A.~E. Myers, et~al.
\newblock Identification of animal movement patterns using tri-axial
  accelerometry.
\newblock {\em Endangered species research}, 10:47--60, 2008.

\bibitem{sherub2016behavioural}
S.~Sherub, G.~Bohrer, M.~Wikelski, and R.~Weinzierl.
\newblock Behavioural adaptations to flight into thin air.
\newblock {\em Biology letters}, 12(10):20160432, 2016.

\bibitem{sherub2017bio}
S.~Sherub, W.~Fiedler, O.~Duriez, and M.~Wikelski.
\newblock Bio-logging, new technologies to study conservation physiology on the
  move: a case study on annual survival of himalayan vultures.
\newblock {\em Journal of Comparative Physiology A}, 203(6-7):531--542, 2017.

\bibitem{spretke2011exploration}
D.~Spretke, P.~Bak, H.~Janetzko, B.~Kranstauber, F.~Mansmann, and S.~Davidson.
\newblock Exploration through enrichment: a visual analytics approach for
  animal movement.
\newblock In {\em Proceedings of the 19th ACM SIGSPATIAL International
  Conference on Advances in Geographic Information Systems}, pp. 421--424. ACM,
  2011.

\bibitem{stein2019movement}
M.~Stein, D.~Seebacher, T.~Karge, T.~Polk, M.~Grossniklaus, and D.~A. Keim.
\newblock From movement to events: Improving soccer match annotations.
\newblock In {\em International Conference on Multimedia Modeling}, pp.
  130--142. Springer, 2019.

\bibitem{swan2006removing}
G.~Swan, V.~Naidoo, R.~Cuthbert, R.~E. Green, D.~J. Pain, D.~Swarup,
  V.~Prakash, M.~Taggart, L.~Bekker, D.~Das, et~al.
\newblock Removing the threat of diclofenac to critically endangered asian
  vultures.
\newblock {\em PLoS biology}, 4(3):e66, 2006.

\bibitem{thiebault2013splitting}
A.~Thiebault and Y.~Tremblay.
\newblock Splitting animal trajectories into fine-scale behaviorally consistent
  movement units: breaking points relate to external stimuli in a foraging
  seabird.
\newblock {\em Behavioral Ecology and Sociobiology}, 67(6):1013--1026, 2013.

\bibitem{truong2018review}
C.~Truong, L.~Oudre, and N.~Vayatis.
\newblock A review of change point detection methods.
\newblock {\em arXiv preprint arXiv:1801.00718}, 2018.

\bibitem{vamocs2017multiscale}
C.~Vamo{\c{s}}.
\newblock Multiscale structure of time series revealed by the monotony
  spectrum.
\newblock {\em Physical Review E}, 95(3):033310, 2017.

\bibitem{walker2015timeclassifier}
J.~S. Walker, M.~W. Jones, R.~S. Laramee, O.~R. Bidder, H.~J. Williams,
  R.~Scott, E.~L.~C. Shepard, and R.~P. Wilson.
\newblock Timeclassifier: a visual analytic system for the classification of
  multi-dimensional time series data.
\newblock {\em The Visual Computer}, 31(6-8):1067--1078, 2015.

\bibitem{yao1988estimating}
Y.-C. Yao.
\newblock Estimating the number of change-points via schwarz'criterion.
\newblock {\em Statistics \& Probability Letters}, 6(3):181--189, 1988.

\bibitem{yeh2017multivariate}
C.-C.~M. Yeh, N.~Kavantzas, and E.~Keogh.
\newblock Matrix profile vi: Meaningful multidimensional motif discovery.
\newblock pp. 565--574, 11 2017. doi: {{%
10\hspace{.1pt}\discretionary{.}{%
}{.}\hspace{.4pt}1109\discretionary{/}{%
}{/}ICDM\hspace{.1pt}\discretionary{.}{%
}{.}\hspace{.4pt}2017\hspace{.1pt}\discretionary{.}{%
}{.}\hspace{.4pt}66}}


\bibitem{yeh2016matrix}
C.-C.~M. Yeh, Y.~Zhu, L.~Ulanova, N.~Begum, Y.~Ding, H.~A. Dau, D.~F. Silva,
  A.~Mueen, and E.~Keogh.
\newblock Matrix profile i: all pairs similarity joins for time series: a
  unifying view that includes motifs, discords and shapelets.
\newblock In {\em Data Mining (ICDM), 2016 IEEE 16th International Conference
  on}, pp. 1317--1322. IEEE, 2016.

\bibitem{zhao2011kronominer}
J.~Zhao, F.~Chevalier, and R.~Balakrishnan.
\newblock Kronominer: using multi-foci navigation for the visual exploration of
  time-series data.
\newblock In {\em Proceedings of the SIGCHI Conference on Human Factors in
  Computing Systems}, pp. 1737--1746, 2011.

\bibitem{zhu2016matrix}
Y.~Zhu, Z.~Zimmerman, N.~S. Senobari, C.-C.~M. Yeh, G.~Funning, A.~Mueen,
  P.~Brisk, and E.~Keogh.
\newblock Matrix profile ii: Exploiting a novel algorithm and gpus to break the
  one hundred million barrier for time series motifs and joins.
\newblock In {\em 2016 IEEE 16th international conference on data mining
  (ICDM)}, pp. 739--748. IEEE, 2016.

\bibitem{zucchini2017hidden}
W.~Zucchini, I.~L. MacDonald, and R.~Langrock.
\newblock {\em Hidden Markov Models for Time Series: An Introduction Using R,
  Second Edition}.
\newblock Chapman \& Hall/CRC Monographs on Statistics \& Applied Probability.
  CRC Press, 2017.

\bibitem{entso}
{\noopsort{zzz1-example}}{Unicorn Systems}.
\newblock Entso-e transparency platform.
\newblock \\\url{https://transparency.entsoe.eu/}.
\newblock Accessed: 2020-04-28.

\bibitem{Icarus}
{\noopsort{zzz2-example}}{Icarus Initiative}.
\newblock Icarus - global monitoring with animals.
\newblock \\\url{https://www.icarus.mpg.de/en}.
\newblock Accessed: 2020-04-11.

\bibitem{Movebank}
{\noopsort{zzz3-example}}{M. Wikelski, S. C. Davidson, and R. Kays}.
\newblock Movebank: archive, analysis and sharing of animal movement data.
  hosted by the max planck institute of animal behavior.
\newblock \url{http://www.movebank.org}.
\newblock Accessed: 2020-04-11.

\bibitem{griffonstudy}
{\noopsort{zzz4-example}}{S. Sherub and M. Wikelski}.
\newblock Movebank: e-obs gprs himalayan griffon, sherub, bhutan.
\newblock
  \url{https://www.movebank.org/cms/webapp?gwt_fragment=page=studies,path=study20202974}.
\newblock Accessed: 2020-04-29.

\end{thebibliography}

\end{document}